\documentclass[amsmath,showpacs,twocolumn,aps,pra,superscriptaddress]{revtex4}

\usepackage{amsmath,amsfonts,amssymb}
\usepackage{graphicx}
\usepackage{color}
\usepackage{enumerate}
\usepackage{bbm}
\usepackage{float}
\usepackage{verbatim}

\makeatletter

\newcommand{\Rmnum}[1]{\expandafter\@slowromancap\romannumeral #1@}
\makeatother

\newcommand{\be}{\begin{equation}}
\newcommand{\ee}{\end{equation}}
\newcommand{\bea}{\begin{eqnarray}}
\newcommand{\eea}{\end{eqnarray}}

\newcommand{\e}{\mathrm{e}}

\date{\today}

\begin{document}

\title{Phase diagram for the Harper model of the honeycomb lattice}

\author{Geo Jose}
\affiliation{Department of Physics, Indian Institute of Science Education and Research, Bhopal, India}

\author{Rajesh Malla}
\affiliation{Department of Physics, Indian Institute of Science Education and Research, Bhopal, India}
\author{Varadharajan Srinivasan}
\affiliation{Department of Chemistry, Indian Institute of Science Education and Research, Bhopal, India}
\affiliation{Department of Physics, Indian Institute of Science Education and Research, Bhopal, India}
\author{Auditya Sharma}
\affiliation{Department of Physics, Indian Institute of Science Education and Research, Bhopal, India}
\author{Suhas Gangadharaiah}
\affiliation{Department of Physics, Indian Institute of Science Education and Research, Bhopal, India}

\begin{abstract}
The Harper equation arising out of a tight-binding model of electrons on a honeycomb lattice subject to a uniform magnetic field perpendicular to the plane is studied. Contrasting and complementary approaches involving von Neumann entropy, fidelity, fidelity susceptibility, multifractal analysis are employed to characterize the phase diagram. The phase diagram consists of three phases: two metallic phases and an insulating phase. A variant model where next nearest neighbor hopping is included, exhibits a mobility edge and does not allow for a simple single phase diagram characterizing all the eigenstates.\end{abstract}


\maketitle

\section{Introduction}
The localization of electronic states in a random medium due to
quantum interference effects is known as Anderson
localization~\cite{Ander}. A consequence of this phenomenon is insulating
behaviour in a one-dimensional (1D) lattice of non-interacting
electrons subject to a random potential~\cite{Elihu}. It is also the
driving mechanism behind the metal-insulator transition (MIT) in
disordered materials in 3D~\cite{Elihu, G4, Imada}.  However, for
quasi-periodic systems such as the Aubry-Andre-Harper model (AAH) 
 it is well known  that
 a metal-insulator phase transition is exhibited even in 1D~\cite{Harper, Aubry}.

The AAH model is obtained from a 1D tight-binding Hamiltonian
and is given by
\begin{eqnarray}
t_{i+1}\phi_{i+1} + t_{i-1}\phi_{i-1} + V_i \phi_{i} =E\phi_{i},
\end{eqnarray}
where $t_i$, $V_i$ and  $\phi_{i}$ are  the 
hopping terms, the on-site potential term 
and the wave function at the $i^{th}$ site, respectively. For the choice $t_i=1$ 
and $V_i = \lambda \cos(2 \pi \phi i + \theta)$ the above equation
 is derivable from the 2D square lattice nearest-neighbor  hopping model  in the
presence of a  uniform magnetic field~\cite{Hofst}. 
For irrational values of  $\phi$, the potential term $V_i$ is quasi-periodic and exhibits a localization-delocalization transition at  $\lambda=2$~\cite{Aubry}.  Below this value all the eigenstates are extended, 
while they are all  localized for $\lambda > 2$. The  wave-functions 
at $\lambda=2$ are critical, exhibiting features such as multi-fractality. 

Generalisations of the above model (extended Harper model) include
next-nearest-neighbor (NNN) hopping on a square lattice and
nearest-neighbour hopping on a two-dimensional triangular lattice both
in the presence of a uniform magnetic field~\cite{Han}.  Interestingly, in both these cases, the phase
diagram consists of one localized phase and two metallic phases
separated by critical lines and one bicritical point where the three
phases meet. Over the years a number of techniques have
been used to characterize the phases.  Analytical techniques include
calculation of the Lyapunov exponent~\cite{Halsey} and measure of the
spectrum~\cite{Thouless}.  Numerical techniques include the study of
level statistics, in particular the distribution of normalised energy
gaps and bandwidths at the critical lines~\cite{Evangelou,Takada,
  Ino}; in addition multifractal analysis of the spectrum and the
wave-function have proved useful in characterising the critical
regions~\cite{Kohmoto1,Kohmoto2}.

\begin{figure}
\includegraphics[width=0.9\columnwidth]{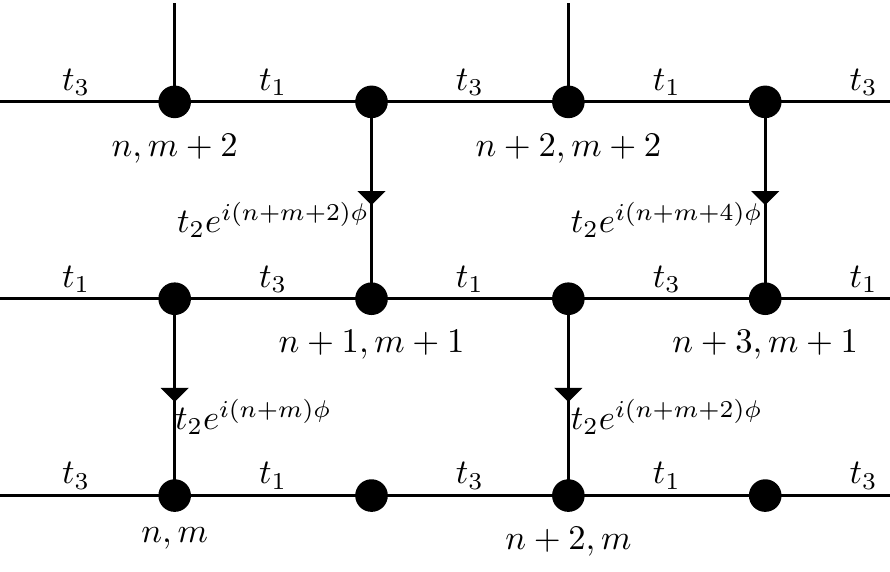}\
\caption{Honeycomb lattice with nearest neighbor hopping in the presence of a uniform magnetic field.}\label{lattice}
\end{figure}

In recent years the application of quantum information techniques in
the context of quantum-phase transition (QPT) of a many-body system
has led to unique insights~\cite{Zanardi, Quan}.  In this regard the
study of quantum fidelity and fidelity susceptibility have been quite
useful in
 a system where the knowledge of 
the relevant order parameter and the changes in symmetries are not known~\cite{Gu}. 
Quantum fidelity (QF) measures  the overlap between two ground states with differing  parameters, whereas fidelity
susceptibility measures the change in QF between two parameters with infinitesimal separation~\cite{Zanardi, Quan, You, Gu}.  Another 
important tool  
is the von Neumann entropy (VNE)
which quantifies  the entropy of entanglement~\cite{Bennett, Steane}.  These techniques have been implemented to successfully characterise the phase diagram of the 
extended Harper model obtained by considering a triangular lattice~\cite{Gong}. 

On the experimental front, tremendous advancement, especially in the ability to 
use ultra cold atoms and photonic lattices for  controlled simulation of many-body states, has created a lot of excitement~\cite{Lewenstein,Bloch}.
Signatures of the localization-delocalization transition in the AAH model were observed almost a decade ago~\cite{Billy, Roati}.
More complicated models are  being simulated with the help of Raman lasers with  inhomogeneous intensity~\cite{Lin, Dalibard}, lattice modulation
 techniques~\cite{Struck1,Struck2} and laser assisted tunneling~\cite{Kolovsky, Creffield}. These techniques provide non-trivial Berry phases 
 which serve as artificial gauge fields.  The latter technique has been 
utilized to generate large tunable artificial magnetic fields  and spatially dependent complex tunneling amplitudes which were successfully  used  to
simulate the 2D square lattice Hofstadter model~\cite{Aidelsburger,Miyake}.  With work  being done on generating artificial honeycomb lattices for atoms
via optical trapping techniques~\cite{polini,  Safaei}, it can be expected that the 2D honeycomb Hofstadter model will be realized soon.

The energy spectrum of the honeycomb lattice with asymmetric hopping has been shown to
exhibit the Hofstadter butterfly pattern~\cite{Gumbs}.  In this work, we obtain the phase diagram 
 of the honeycomb  lattice with asymmetric nearest neighbour   hopping  in the presence of a uniform magnetic field.  With the help of 
 quantum-information  techniques, we show that similar to  the case of the triangular lattice (and square lattice with NNN hopping),  the phase 
 diagram is divided into 
an insulating region and two metallic regions. However, in contrast to the other two cases, the phase diagram is different in structure.  
Further, multi-fractal analysis and level statistics approaches allow us to show that the lines 
separating the  three regions are critical.  In the last part of this work we  include  NNN  hopping; this results in the appearance of mobility edges.

The organization of this paper is as follows. The next section describes the model, and reports the phase diagram obtained. The third section provides details of the quantum information techniques used to arrive at the phase diagram. The fourth and fifth sections are devoted to multi-fractal analysis and level statistics approaches respectively, for characterizing the critical lines.  This is followed by a section on the NNN hopping model, and a final section of summary and conclusions.

\section{Model }\label{model}

The tight binding Hamiltonian of an electron in a honeycomb lattice in
the presence of a uniform magnetic field has the form:
\begin{widetext}
\begin{equation}
H=- \sum_{n+m=\text{even}} \big[ t_1 a^\dagger_{n+1,m}a_{n,m} + t_2 \e^{-i (n+m) \phi} a^\dagger_{n,m+1}a_{n,m}   
 ++    t_3 a^\dagger_{n-1,m}a_{n,m}  \big] + h.c.,
\end{equation}
\end{widetext}
where $a^\dagger_{n,m}$ ($a_{n,m}$) are the creation (annihilation) operators at the $(n,m)^{\text{th}}$ site, 
  $t_1$, $t_2$ and $t_3$ are real hopping parameters,  the gauge choice is $[0,\phi (x+y),0 ]$,   and   $2\phi$ is the flux enclosed per 
 unit cell [Fig~(\ref{lattice})].  For $\phi=p/q$, where $p$ and $q$ are coprimes,  the
 phase term repeats itself  after $q$ ($2q$) sites along the $x-$direction for $q$ even  (odd).  
 The phase factor  is attached with the $t_2$ hopping term and for fixed $n+m$  remains invariant  for hopping from  $(n,m)\rightarrow (n,m+1)$ site, consequently, the wave-function amplitude at the $(n,m)^{th}$ site can 
 be written as $\Psi_{n,m}=\Psi_{n+m} e^{i(k_x n +k_y m)}$.  
 Therefore the Harper matrix   obtained from the  
Schr\"{o}dinger equation,  $\tilde{H} |\varepsilon, k_x,k_y \rangle = \epsilon |\varepsilon, k_x,k_y \rangle$, where $|\varepsilon, k_x,k_y \rangle = \sum \Psi_{n+m} e^{i(k_x n +k_y m)}  |n,m\rangle$, 
   acquires the following form~\cite{Hasegawa}
 \begin{equation}
\begin{pmatrix}
-E &~ B_1&~ 0 & \cdots &~0& B_m^{'*}\\ \\
B_1^{*}&~ -E&~B_1'&\cdots&~0 &0\\ \\
\vdots &~ \ddots &~ \ddots &~\cdots &~\vdots&\vdots \\ \\
 0 &~ 0 &~ \cdots &~0&~-E& B_m \\ \\
B_m^{'}&~ 0 &~ \cdots &~0&~ B_m^{*} & -E
\end{pmatrix}=0,
\end{equation} 
 where $m=q/2$ for $q$ even, $m=q$ for $q$ odd, $B_n =-t_1 e^{ik_x} -t_2 e^{ik_y} e^{i 4\pi n \phi} $
 and $B_n^{'} =-t_3 e^{ik_x}$.

  The corresponding 
  1D Harper Hamiltonian has the form of a non-interacting chain with asymmetric nearest-neighbor hopping:
 \begin{eqnarray}\label{1DHamil}
&&\tilde{H}= \sum_{s}  \Gamma_s^{l} c_{2s-1}^\dagger c_{2s} +  \Gamma_s^{r}   c^\dagger_{2s+1} c_{2s}+ h.c.,
\end{eqnarray}  
where the left and the right hopping parameters are $\Gamma_s^{l} = -t_1 \e^{ik_x} -t_2 \e^{ik_y + i4\pi s \phi}$ and 
 $\Gamma_s^{r} =-t_3 \e^{-ik_x}$, respectively.  Invoking the Schr\"{o}dinger equation,  $\tilde{H} |\psi\rangle = \epsilon  |\psi\rangle $, where $|\psi\rangle = \sum \psi_s c_s^\dagger  |0\rangle $, the following  eigenfunction equations are obtained:
 \begin{eqnarray}\label{efunction}
\epsilon \psi_{2 s}& =& \Gamma_{s}^{l^*}   \psi_{2s-1} +  \Gamma_{s}^{r*}   \psi_{2  s+1} \nonumber\\
\epsilon \psi_{2s+1}& =& \Gamma_{s+1}^{l}   \psi_{2s+2} +  \Gamma_{s}^{r}  \psi_{2s} 
 \end{eqnarray} 
We re-express    Eq.~\ref{efunction} in terms of the parameters $\mu = t_1/t_3$ and $\lambda= t_2/t_3$ and study the 
phase diagram as a function of $\mu$ vs $\lambda$ for irrational 
values of $\phi$.    We fix   $\phi$  to be the inverse of the golden ratio which we approximate  as $\phi_n=F_{n-1}/F_{n}$,  where $F_n$ is the $n$th number of the Fibonacci series.
\begin{figure}
\includegraphics[width=0.6\columnwidth]{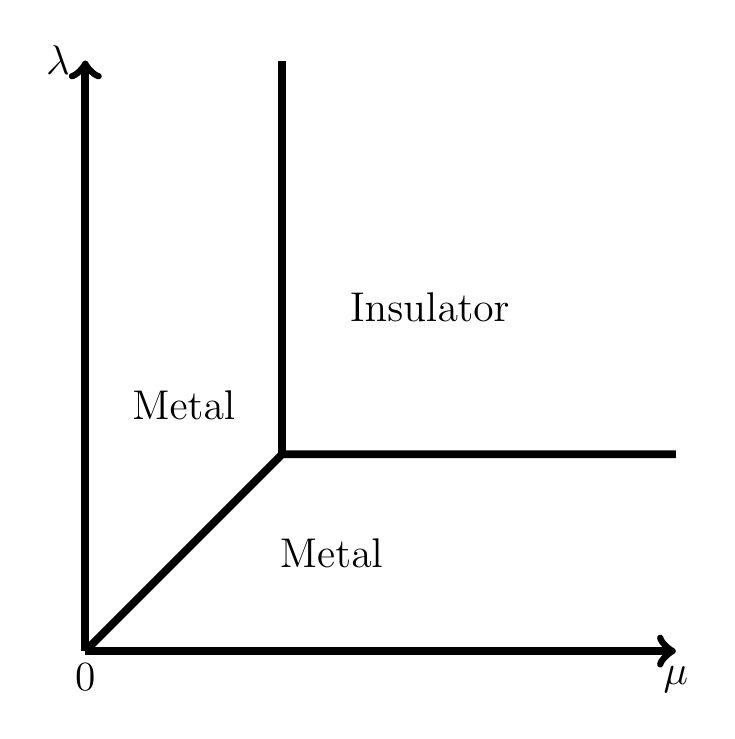}
\caption{The phase diagram as a function of $\mu$ vs $\lambda$. In the regions~I and ~II  the wavefunctions are all delocalized, whereas 
all states are localized in region~III.  The phases are separated by critical lines $\mu=\lambda$, $\mu=1$ and $\lambda=1$ which   meet at $(1,1)$.}\label{phase}
\end{figure}
We find that the phase diagram, as shown in Fig~(\ref{phase}), 
can  be divided into three distinct regions. Each of the three phases is separated from 
the other by a critical line. These lines meet at the bicritical point, which corresponds to all the tunnelling parameters being identical. 
To characterise these regions we have used multiple techniques.  While VNE, fidelity and fidelity susceptibility are able to characterise  the regions  into insulating or metallic type, we find that multi-fractal analysis and  level statistics approaches are especially helpful
for studying the  wavefunction and the spectral properties at the critical lines.

 We would like to add that unlike the various versions of the AAH model which have a quasiperiodic  on-site potential term, 
 our effective 1D model can be thought of as a modified  version of the Su-Schrieffer-Heeger model
  with a modulating hopping term.  In this work we have focussed on the ground state as well as bulk state properties of the wave-function to obtain the phase diagram and not  on the edge states. The study of edge states  by    characterizing the  gap via the calculation of the bulk Chern number or by counting the number of  states 
 between the band gap  reveal the  topological properties of  the  
 model. For a given set of parameters, the topological features undergo drastic modification with change in $\phi$~\cite{Gumbs}.  However, the phase diagram we discuss are robust to changes in $\phi$ as long as it can be approximated as an irrational number.  
 
 \section{Phase diagram}
\subsection{von Neumann Entropy}
\begin{figure}
a)\includegraphics[width=0.6\columnwidth]{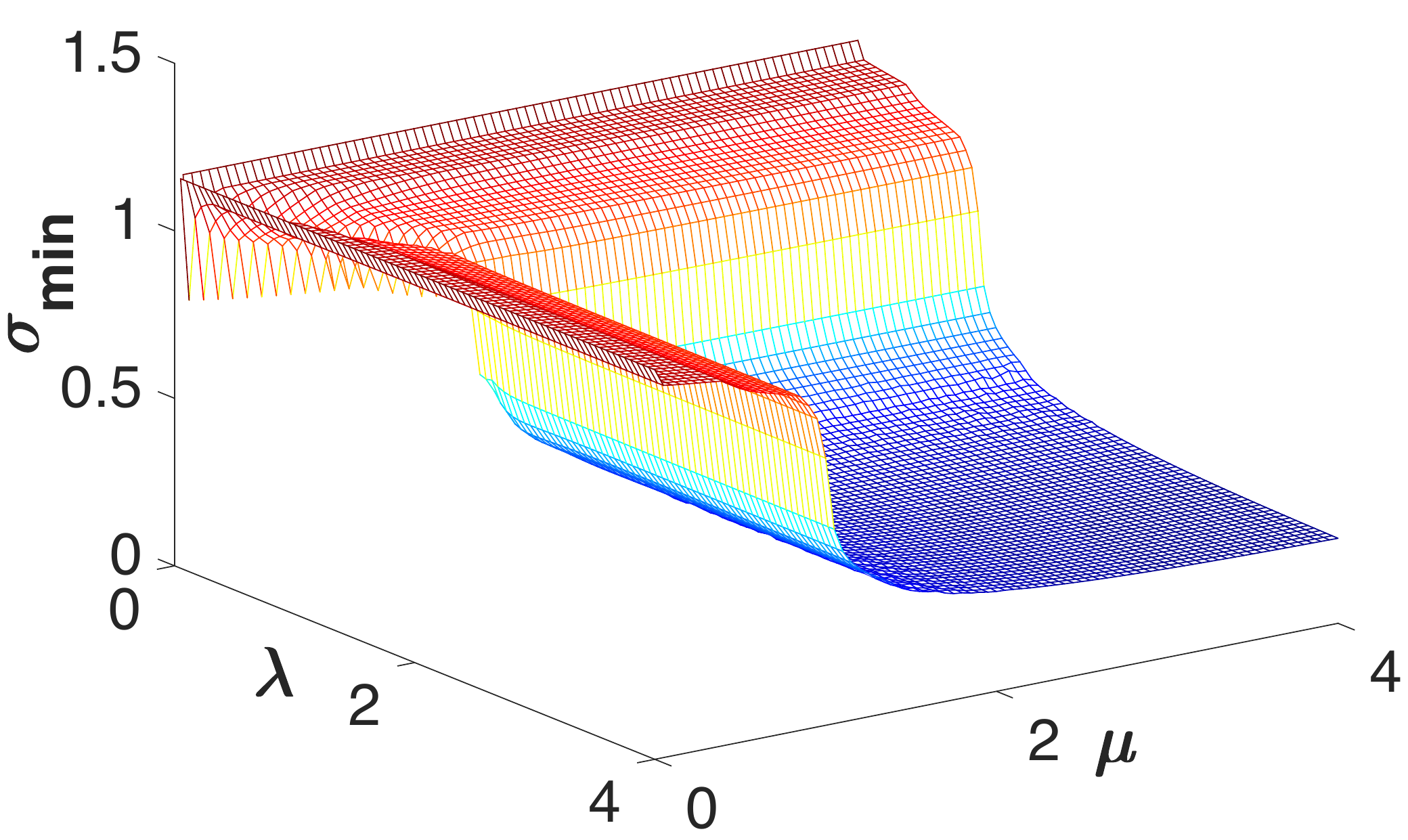}\\
b)\includegraphics[width=0.6\columnwidth]{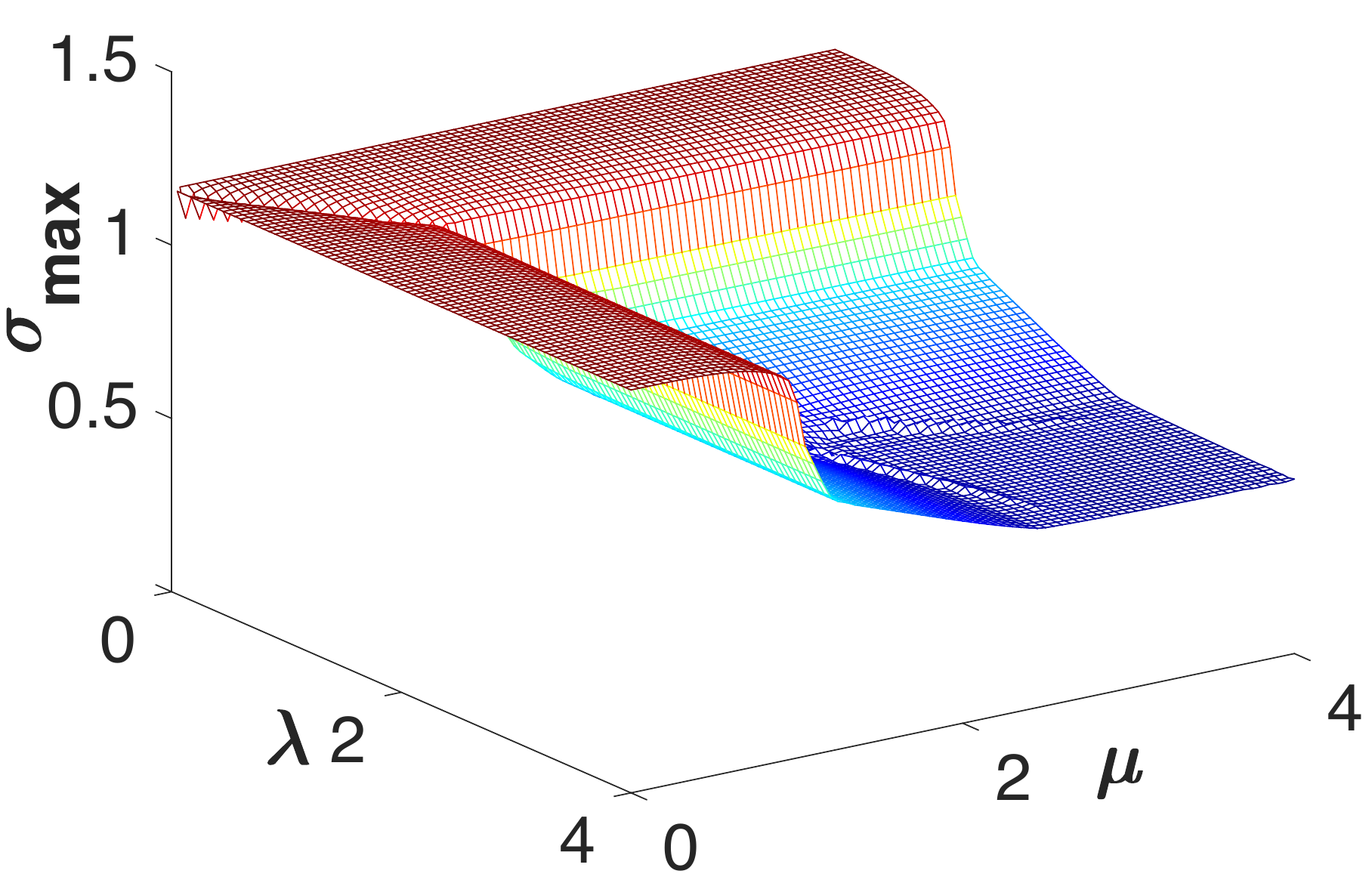}
\caption{Scaled entropy (Eq.~\ref{eqn:sigma}) plots for $F_{n=12}$ as a function of the  parameters $(\mu, \lambda)$ a) minimum  and b) maximum entropy.}\label{fig_entropy}
\end{figure}

We begin our analysis by studying the   von Neumann entropy of  the N-site wave function averaged over all sites~\cite{Gong}:
\begin{eqnarray}
S_{j} = -\sum_{i=1}^{N}\frac{p^{i}_{j}\log_2 p^{i}_{j}   +(1-p^{i}_{j}) \log_2(1- p^{i}_{j})      }{N},
\end{eqnarray} 
where $p^{i}_{j}$ is the occupation probability of the $i^{\text{th}}$ site  of the $j^{\text{th}}$ wave-function. 
We note that for delocalised wave functions  $p^{i}_{j}\sim 1/N$
 and hence $S_{j} \sim \log_2(N) / N $. On the other hand,  $S_{j}  \sim 1/N$  for wave-functions that are localised.
Thus  the von Neumann entropy  serves as an excellent diagnostic tool for distinguishing regions that are localised from  those
that are delocalised.

It is useful to define the scaled entropies as:
 \begin{equation}
 \label{eqn:sigma}
  \sigma_{j} =  \frac{N} {\log_2 N}S_{j}.
 \end{equation}
In  Figs.~\ref{fig_entropy} $(a)$ and $(b)$, we plot  the minimum and maximum scaled entropy from the  set $ \{\sigma_1,...,\sigma_N\}$ at each parameter value $( \mu,\lambda)$.  Interestingly, even for relatively small system sizes  (for example $F_{n=12}$),  
a clear separation between the extended regions~II,~III and the localized region~I, is seen.  Another feature that is illustrated from the near identical phase diagrams of Figs.~\ref{fig_entropy}  $(a)$ and $(b)$ is that there are no mobility edges.
The wave functions in a given parameter regime are either all localised or all delocalised.

\subsection{Fidelity and Fidelity Susceptibility}
Although the VNE  approach is able to  distinguish
regions into metallic and insulating type,   the distinction between the two metallic regions remain inconclusive.
We find that fidelity and FS techniques are able to further characterise the metallic regions.

\begin{figure}
a)\includegraphics[width=0.46\columnwidth]{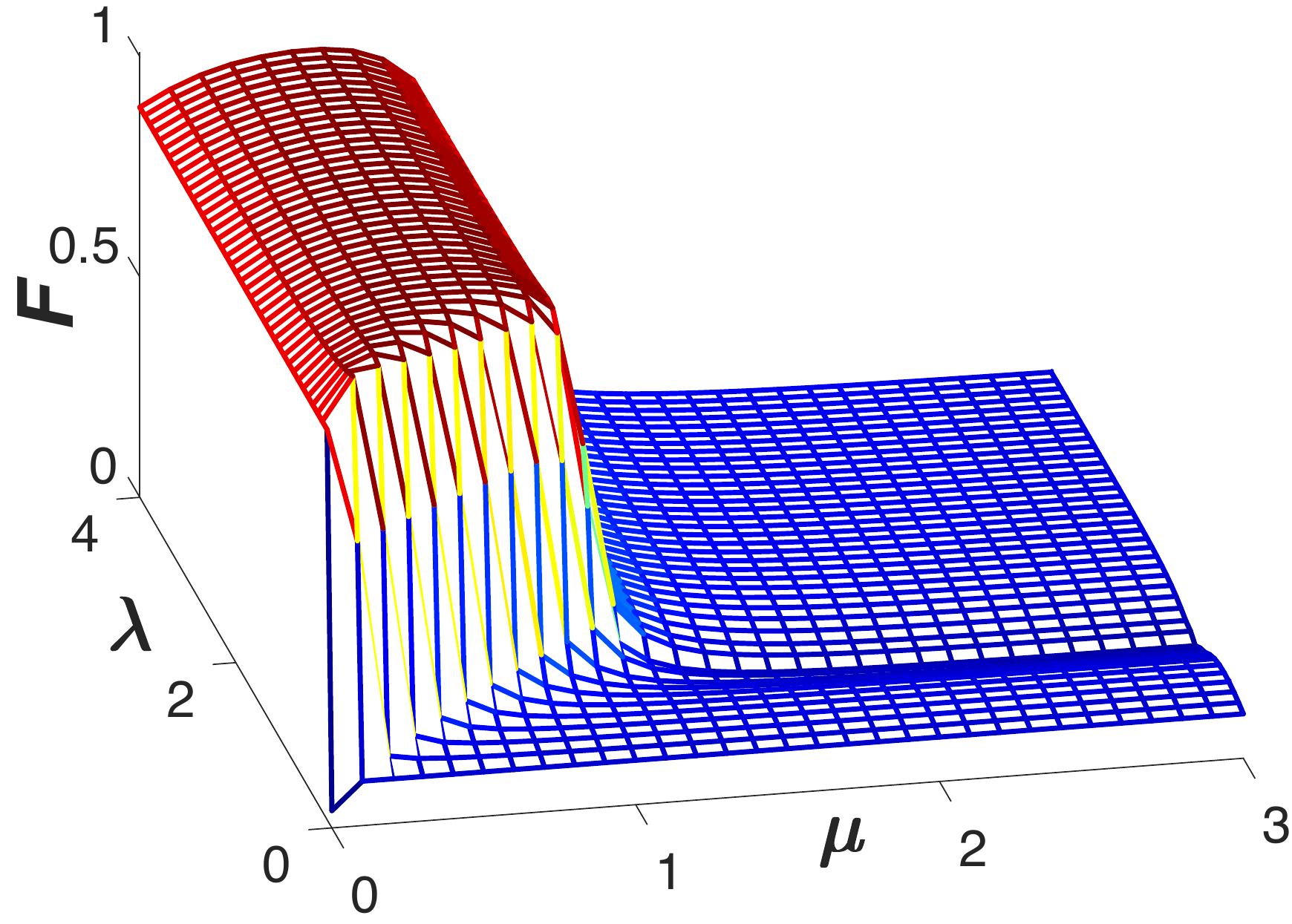}\
b)\includegraphics[width=0.46\columnwidth]{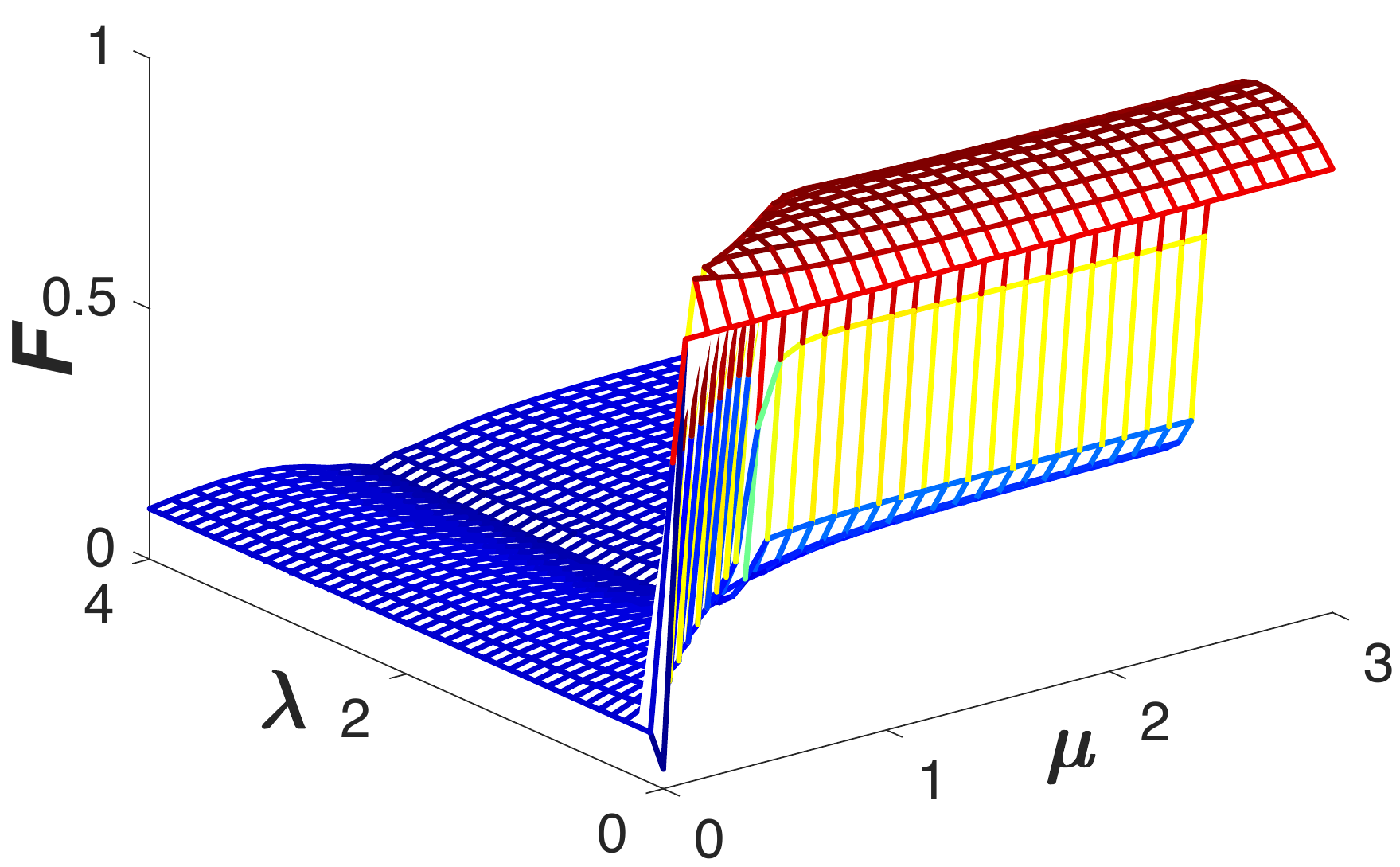}\
c)\includegraphics[width=0.46\columnwidth]{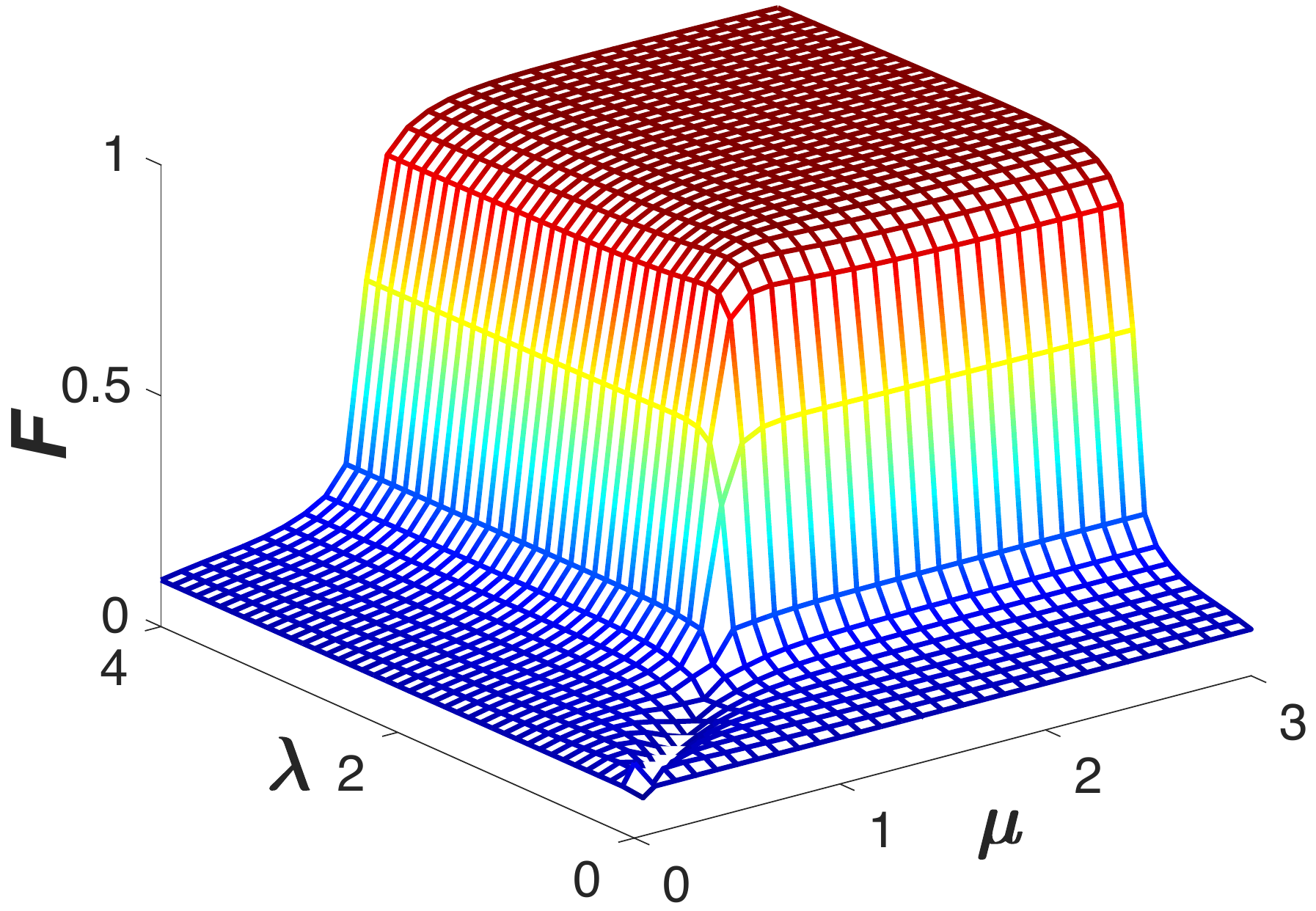}\
d)\includegraphics[width=0.46\columnwidth]{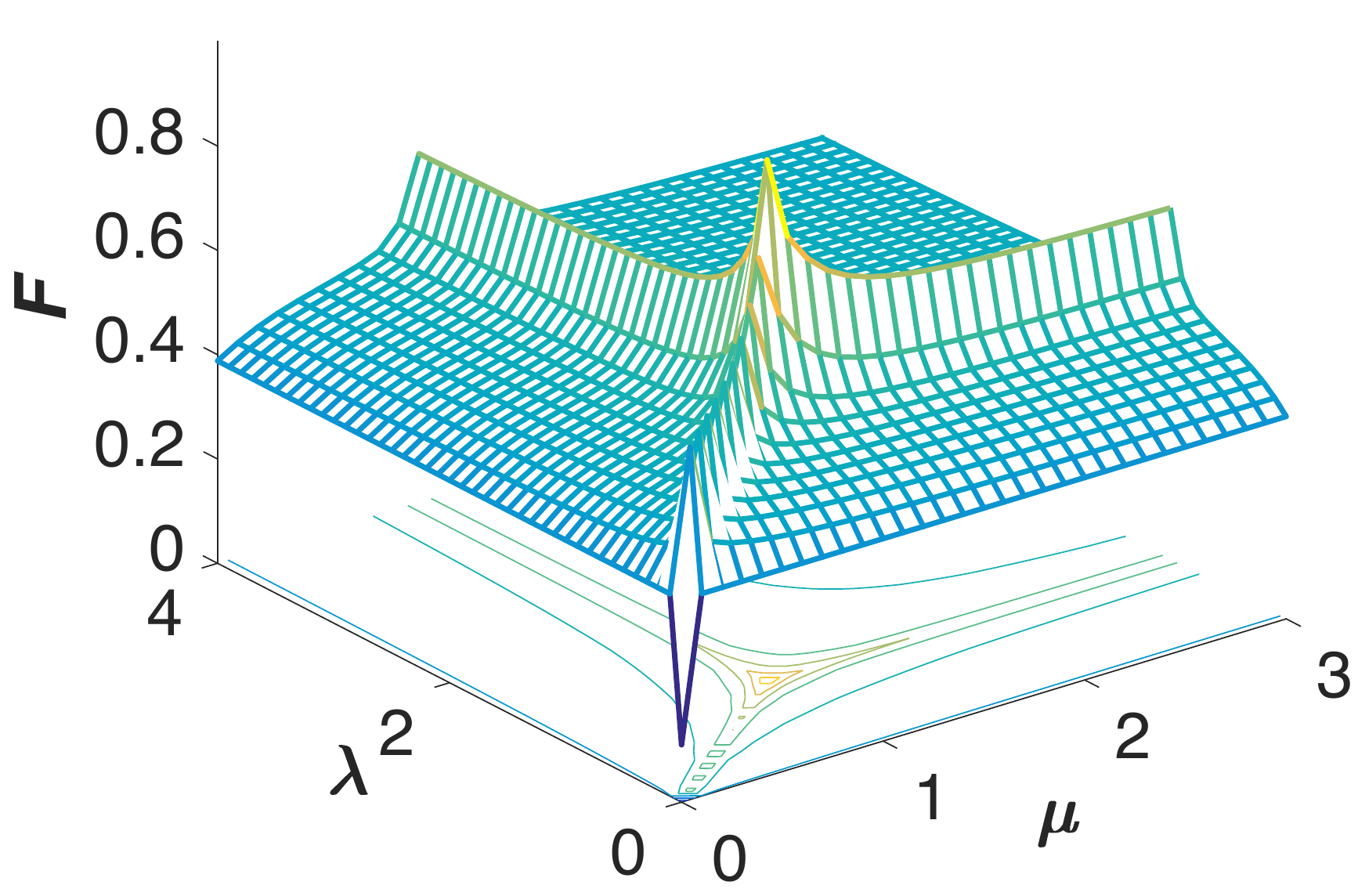}
\caption{Fidelity plots for $p/q = 144/233$ with the  reference paramters $(\mu, \lambda)$: a) $(0.5,1)$, b) $ (1,0.5)$, which are   in the metallic regions $1$ and $2$, respectively,  c) $(2,2)$ in the insulating regime $3$ and  d) the reference point  $(1,1)$ which is the bicritical point.}\label{fig_fidelity}
\end{figure}

Fidelity is defined as the overlap of the  ground-state wave-functions  for  differing  parameters and is given by  
\begin{equation}
F(\Upsilon_1,\Upsilon_2) =|\langle \psi_0(\Upsilon_1)|   \psi_0(\Upsilon_2) \rangle|^2,
\end{equation}
 where $|   \psi_0 (\Upsilon_{i}) \rangle$  are the ground state wave-functions with
parameters $\Upsilon_{i} = (\mu_{i}, \lambda_{i}), i\in \{1,2\}$.  Keeping one of the parameters as the
reference point, say $\Upsilon_1$, the other parameter  $\Upsilon_2$, is varied. If the two parameters
lie in the same phase the fidelity value is near unity. On the other
hand, the ground-state wave-function differs significantly  for parameters  separated by a quantum critical point/line
thus resulting in near vanishing fidelity values.   In Figs.~\ref{fig_fidelity} (a), (b) and  (c), 
we choose $(0.5,1)$,  $(1,0.5)$ and  $(2,2)$ respectively, as the reference parameters. These highlight,  respectively,   insulating 
and the two metallic regions. Although the data presented is for the lowest energy state,   
 we have checked that the same phase diagram is generated if arbitrary eigenstates are considered.
The two metallic phases represent open orbits along different 
directions in the phase space of the underlying `classical-Hamiltonian'~\cite{Ino}.    An alternate insight into
the various phases is  obtained by studying limiting scenarios in Eqn.~\ref{1DHamil}. For 
$\lambda=0$, the 1D Hamiltonian is metallic with unit-cell made of just two  sites, whereas for the $\mu=0$ case, 
the unit cell is $\propto p$-sites and thus corresponds to an altogether different metallic phase. The $\mu \gg 1$  and $\lambda \gg 1$ regime is insulating since it is  equivalent to  cutting off of $right-$hopping from  an even-site to an odd-site,
thus generating localized islands.

\begin{figure}
\includegraphics[width=0.86\columnwidth]{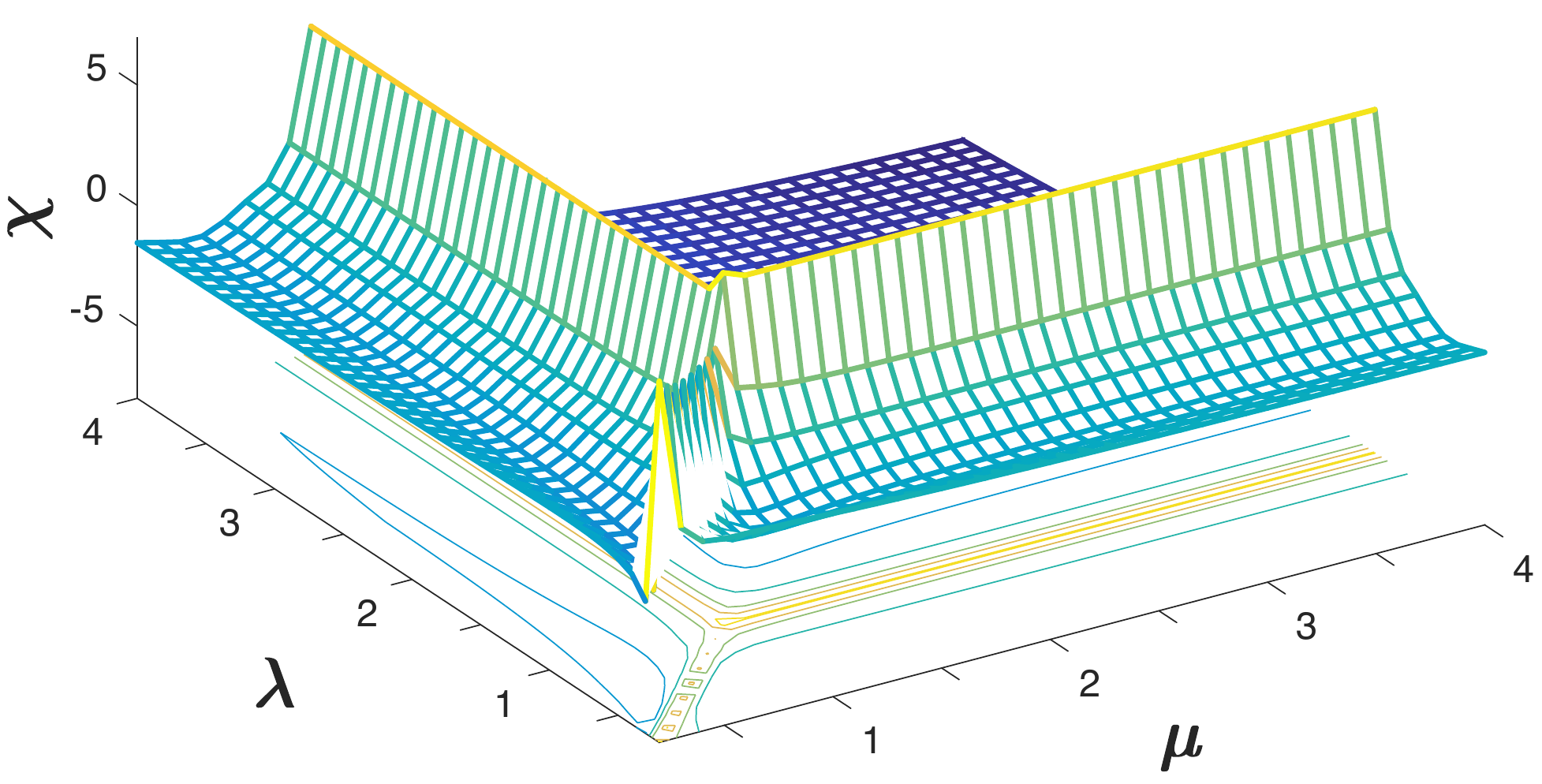}\
\caption{Fidelity-susceptibility as a function of $(\mu,\lambda)$. The unit vector is taken to be  along $\hat{n} =(1/\sqrt{3},\sqrt{2/3})$.}\label{fig_susceptibility}
\end{figure}

Fidelity measurements can also highlight the critical lines. 
When the  reference point is taken to be on a critical line, fidelity exhibits
its maximum value along that line and a sharp decline away from it.  This is best illustrated 
for the reference point $(1,1)$ (which is a bicritical point) for which fidelity exhibits peak behaviour [Figs.~\ref{fig_fidelity}  d]  along all the critical lines. The above approach to isolate the critical lines, is dependent on the correct positioning of the reference point. 
It turns out that an efficient alternate way is to use fidelity susceptibility where no guess work is required. 
Fidelity susceptibility is obtained from the 
fidelity $F(\Upsilon , \Upsilon+\delta \Upsilon) $  between the 
ground states of adjacent parameters~\cite{Zanardi, Quan, You, Gu, Gong}:
\begin{widetext}
\begin{eqnarray}
\chi_F = \sum_{(a,b)=\lambda,\mu}\hat{n}_a \hat{n}_b \big[\frac{1}{2}(\langle \partial_a \psi_0 | \partial_b \psi_0 \rangle   +\langle \partial_b \psi_0 | \partial_a \psi_0 \rangle   ) -\langle \partial_a \psi_0 | \psi_0 \rangle \langle \psi_0 | \partial_b \psi_0 \rangle \big],
\end{eqnarray}
\end{widetext}
where    the   partial 
derivatives are in the parameter space and the unit vector  
$\hat{n}$ represents the direction connecting the two infinitesimally close
points.  The choice of the unit vector is  arbitrary. If the unit vector 
is chosen to be parallel to one of the  critical lines, then all  critical lines except for the one parallel to the unit vector will be exposed.
In our case, fidelity susceptibility plots obtained by considering atleast two choices of non-collinear
unit vectors will reveal all the critical lines.  In Fig.~\ref{fig_susceptibility} we consider $(1/\sqrt{3},\sqrt{2/3})$ as the   the unit vector; 
since this vector is not along any of the critical lines, only one plot is sufficient to  highlight the position and directions   of all the critical lines.

\subsection{Multifractal analysis}
\begin{figure}
a)\includegraphics[width=0.44\columnwidth]{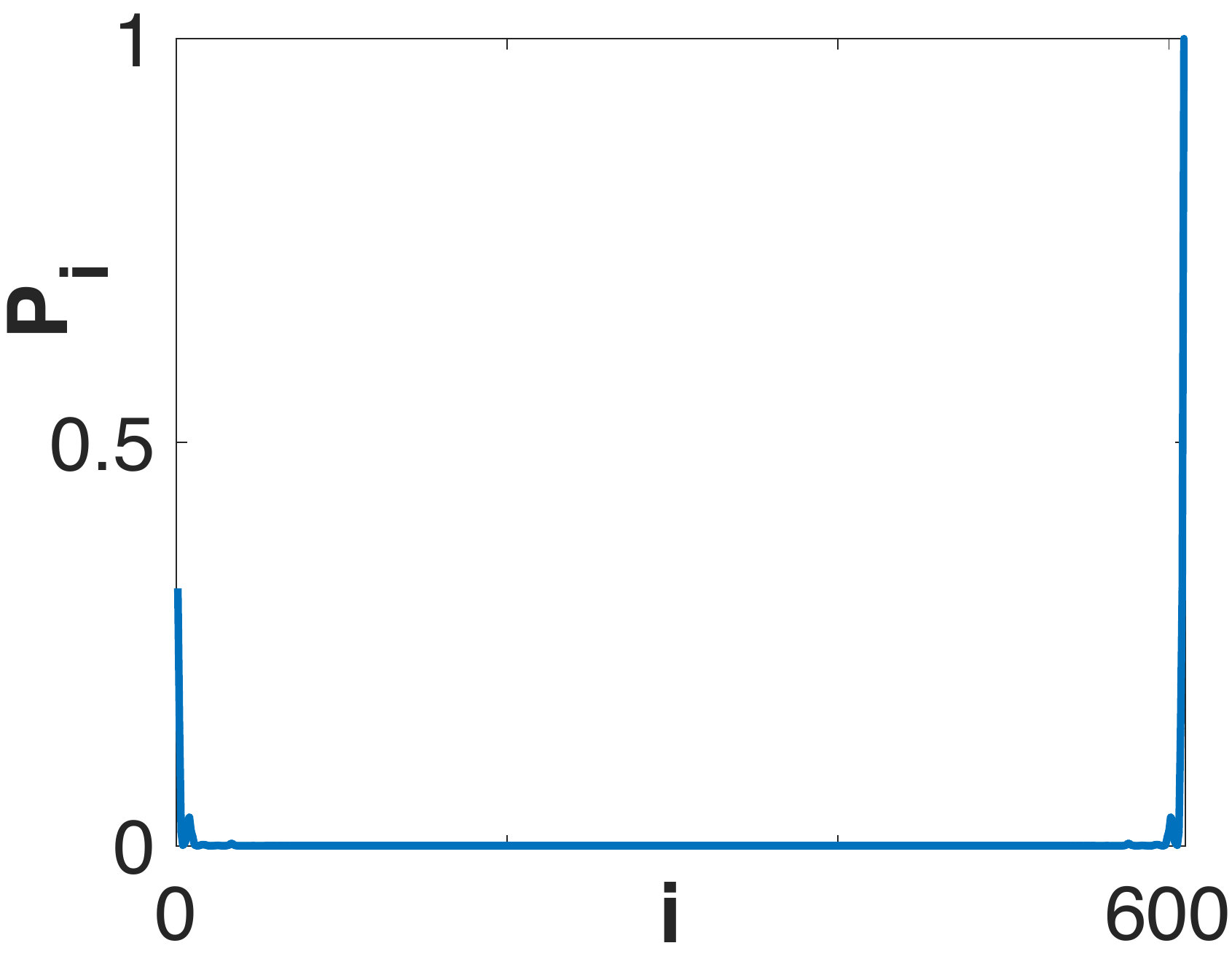}
b)\includegraphics[width=0.46\columnwidth]{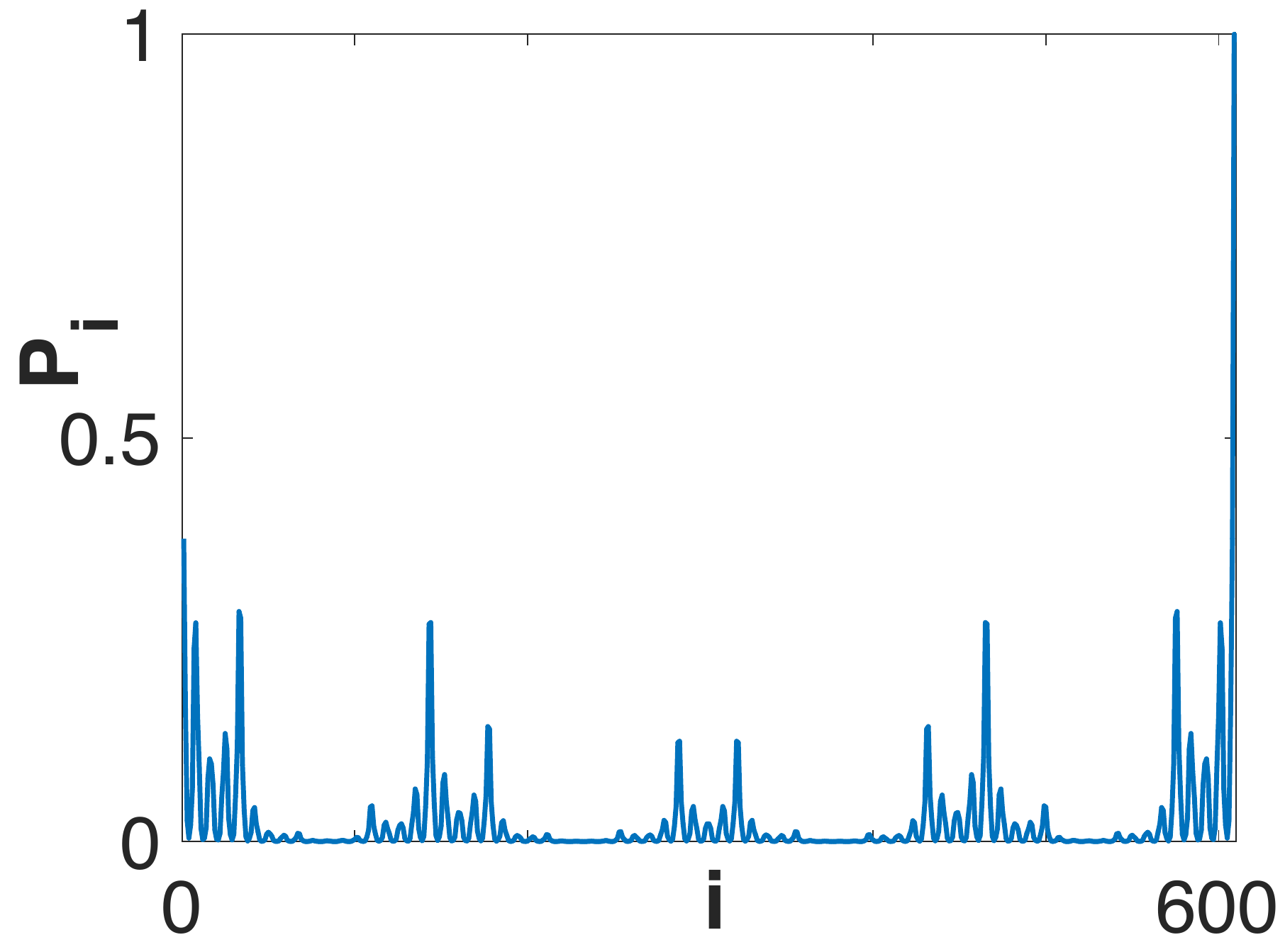}
c)\includegraphics[width=0.56\columnwidth]{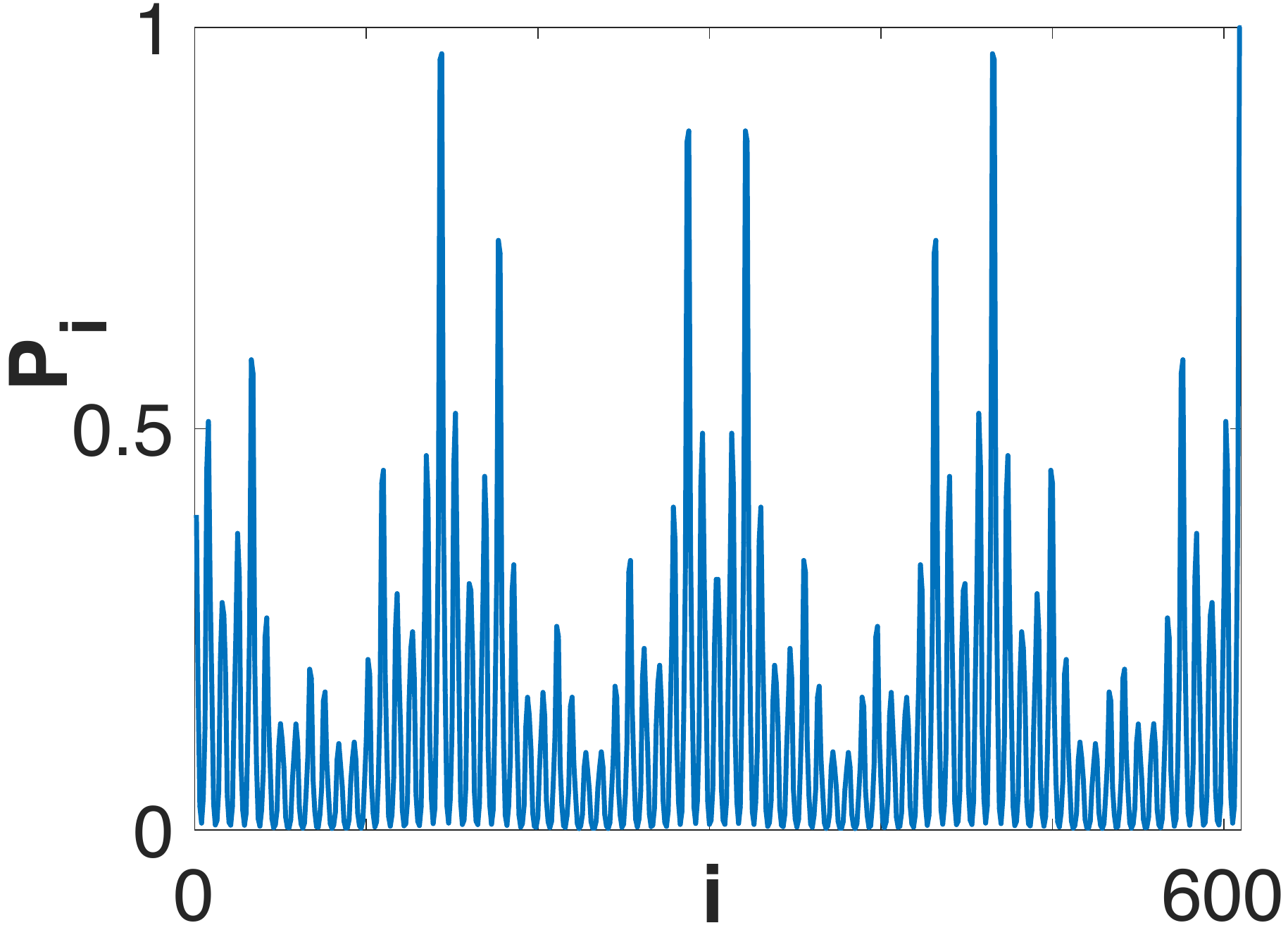}\\
\caption{Ground state wave-function $P_i =\frac{|\psi_i|^2}{|\psi_{\text{max}}|^2}$ for $n=15$, a) in the insulating region (1.1,1.1), b) at the bicritical point (1,1) and, c) the metallic region (0.9,1). }\label{fig_States}
\end{figure}
In this section we will use multifractal  techniques for analysing  the scaling properties of wave functions in 
different regions of  parameter space.  The basic idea is to  express the probability measure of a wave-function at the $i^{th}$ site,  $p_i$, in terms of a
scaling index $\alpha_i$, such that $p_i = N^{-\alpha_{i}}$, where $N\propto F_n$ is the number of lattice sites at the $n^{th}$ step. 
The distribution of $\alpha_{i}$'s thus obtained can be organised in terms of the function $\Omega_n (\alpha)$ defined as the density of sites with scaling index $\alpha$, 
satisfying  $\int \Omega_n (\alpha) d\alpha=N$. Further, $\Omega_n (\alpha)$ is expressed in terms of the scaling function $f_n(\alpha)$
such that $\Omega_n (\alpha) = N^{f_n(\alpha)}$. 

\begin{figure}
\includegraphics[width=0.96\columnwidth]{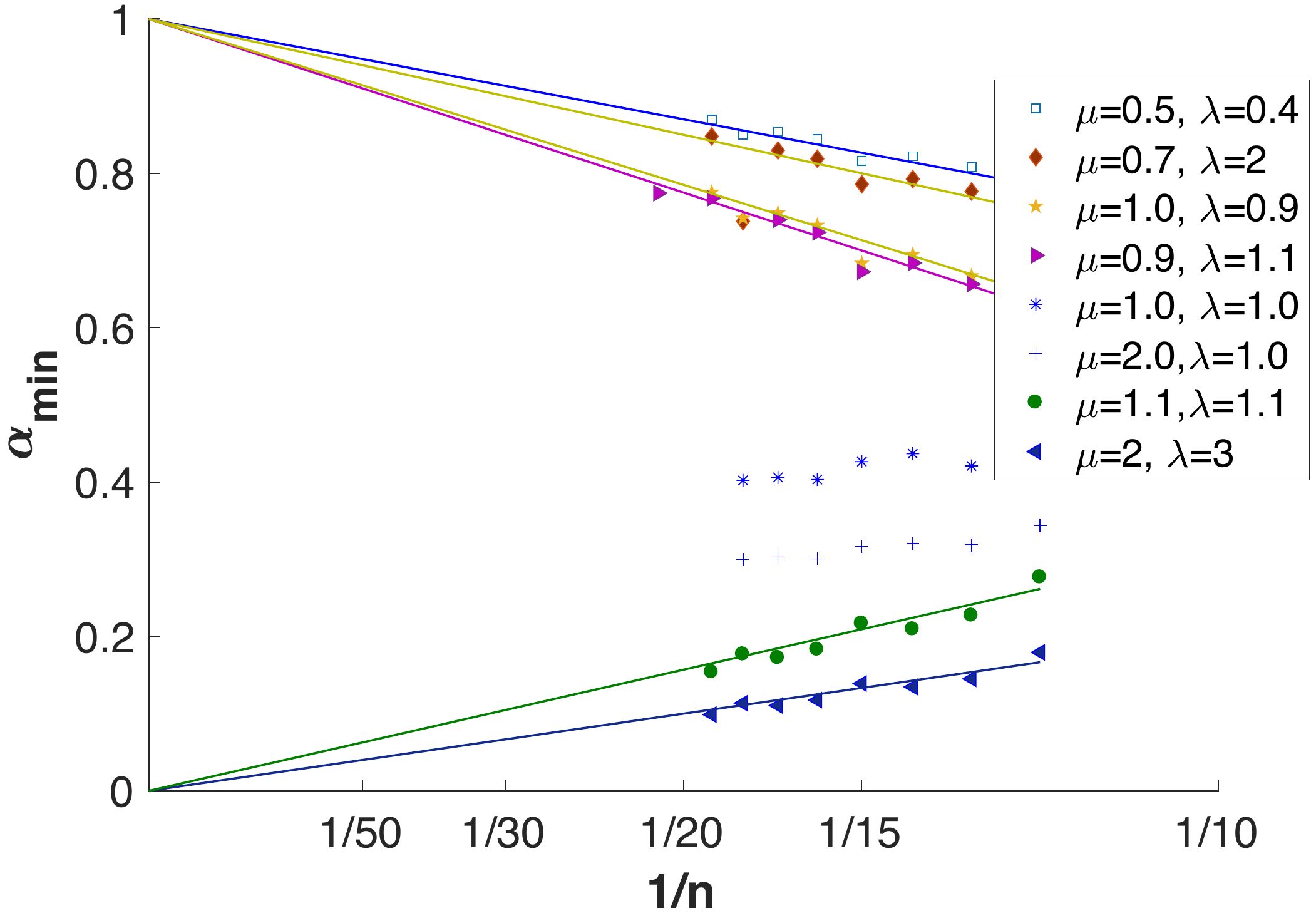}\
\caption{$\alpha_\text{min}$ vs $1/n$. As $n\rightarrow \infty$, $\alpha_\text{min}$ acquires the limiting values $1$ and $0$ for metallic and insulating states, respectively. For critical states  $\alpha_\text{min}$ approaches a value in between the two limits. }\label{fig_MultiFrac}
\end{figure}
 
With the increase in  step-size $n$, the minimum of the scaling index  $\alpha_\text{min}$ 
approaches distinct limits for wave-functions that are extended, localised and critical
(for illustration see Fig.~(\ref{fig_States})).  
In particular, when the states are extended we expect that the probability measure 
at a typical point scale as $p_i \sim1/N$ and consequently $\alpha_\text{min}\rightarrow 1$. On the other hand, the probability measure 
for localised states are concentrated at only few points where the measure is finite. Therefore for localised states  $\alpha_\text{min}\rightarrow 0$, whereas the points
that are unoccupied result in  $\alpha_\text{max}\rightarrow \infty$. 
 Similar arguments for the density of states suggest
that for an extended state $f(\alpha_\text{min})\rightarrow 1$, while for a localised state  $f(\alpha_\text{min})\rightarrow 0$
and $f(\alpha_\text{max})\rightarrow 1$. A critical state is characterised by $0<\alpha_\text{min}<1$ and $f(\alpha)$ defined between the intervals $[\alpha_\text{min},\alpha_\text{max}]$.

\begin{figure}
a)\includegraphics[width=0.46\columnwidth]{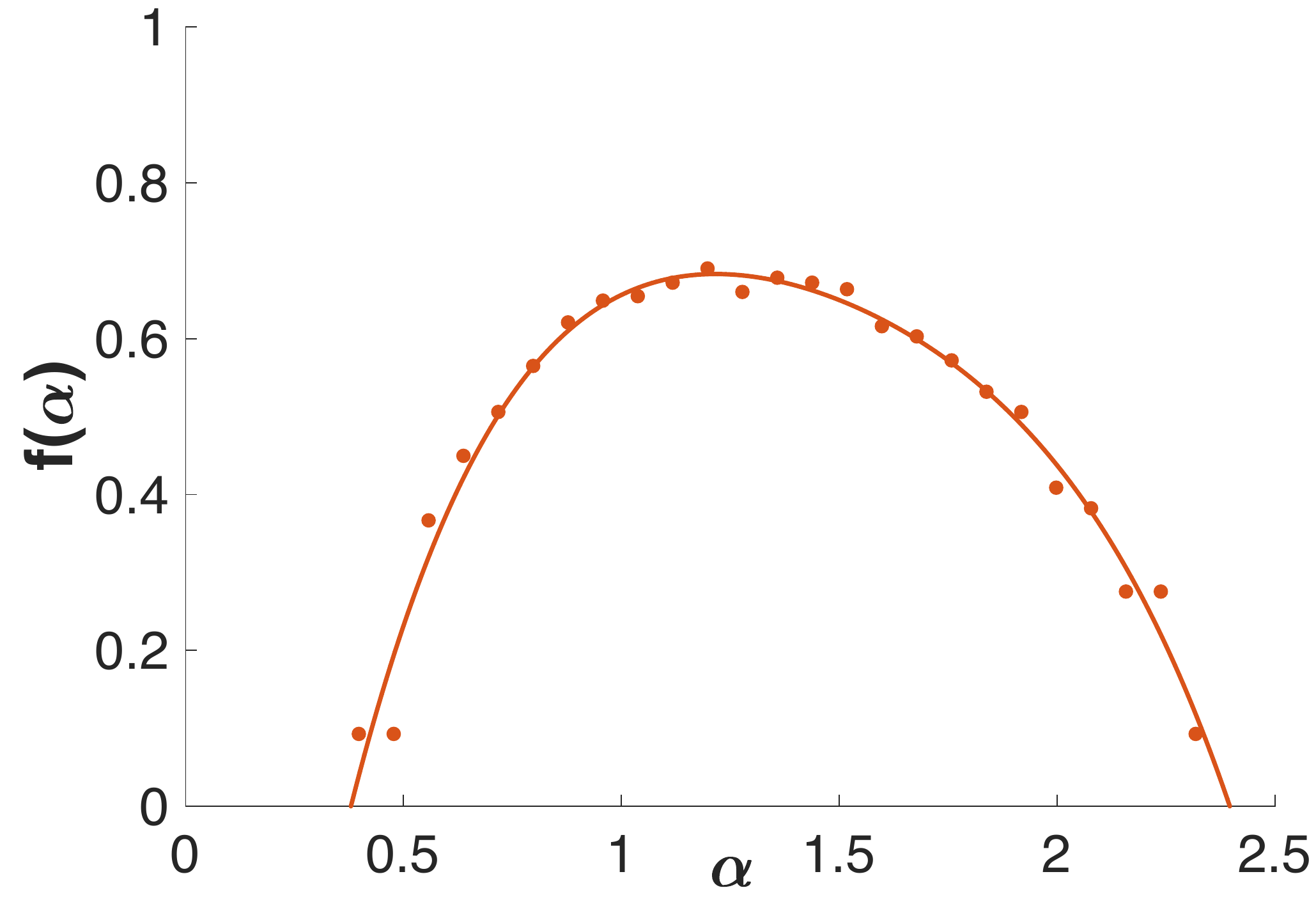}
b)\includegraphics[width=0.46\columnwidth]{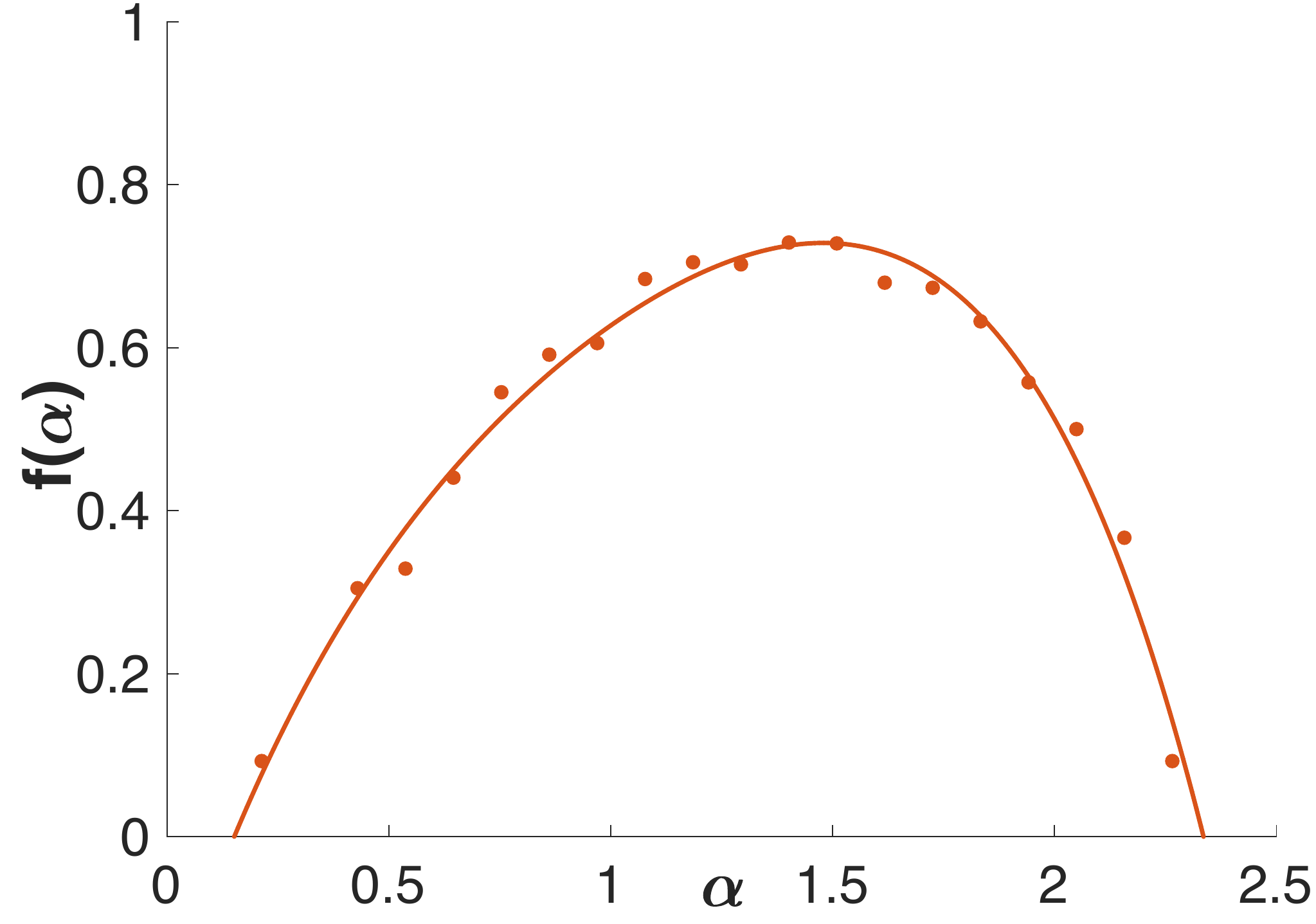}
\caption{$f(\alpha)$ vs $\alpha$ for $n=18$ at the critical points, a) (1,1) and, b) (2,1).  }\label{fig_F_alpha}
\end{figure}

We consider the wave-functions at the lowest end of the spectra and
numerically evaluate  $\alpha_\text{min}$ for Fibonacci indices $n$
and extrapolate them as $n\rightarrow\infty$. The plots of $\alpha_\text{min}$ vs $1/n$  are obtained for parameter points on the critical lines and away from it (see Fig.~\ref{fig_MultiFrac}).
 The plots of $\alpha_\text{min}$ vs $1/n$ for  the parameter points $(\mu,\lambda)=(0.5,0.4),(1,0.9)$ from metallic region I
and $(\mu,\lambda)=(0.7,2),(0,9,1.1)$ from metallic  region II, exhibit $\alpha_\text{min}\rightarrow 1$ as $n\rightarrow \infty$ a behaviour consistent with extended states.  
For $(\mu,\lambda)=(1.1,1.1), (2,3))$ which are in region III, $\alpha_{min}$ 
approaches $0$ as $n\rightarrow\infty$, which is what we expect for an insulating phase.
Finally, we consider   the behavior at the critical points $(\mu,\lambda)=(1,1)$ and  $(2,1)$; we see from Fig.~\ref{fig_MultiFrac} that in the $n\rightarrow\infty$ limit, the exponent $\alpha_\text{min}$  approaches a limit 
different from the metallic and insulating ones and lies between $ 0<\alpha_\text{min}<1$. At the same time as seen in Fig.~\ref{fig_F_alpha},
$f(\alpha)$ acquires a dome like feature for the allowed values of $\alpha$ thus confirming the  multifractal nature at the critical points~\cite{Ino}.

\subsection{Level statistics}
As an alternative quantitative test for the universal features, we investigate the level statistics along the critical lines by studying the behavior of  
the spectral distribution function $P(\omega)$ as $\omega\rightarrow 0$. The normalizations used to define the distribution function  are 
$\int_0^{\infty} P(\omega) d\omega =1$ and  $\int_0^{\infty} \omega P(\omega) d\omega =\langle \omega\rangle=1$.  We find that along 
the critical lines $P(\omega)$ diverges as $\omega^{-\delta}$ ($\delta>0$), a behaviour analogous to earlier studies on the gap distribution 
functions along critical lines for 1D quasi-periodic systems~\cite{Ino,naka}.

\begin{figure}[H]
a)\includegraphics[width=0.46\columnwidth]{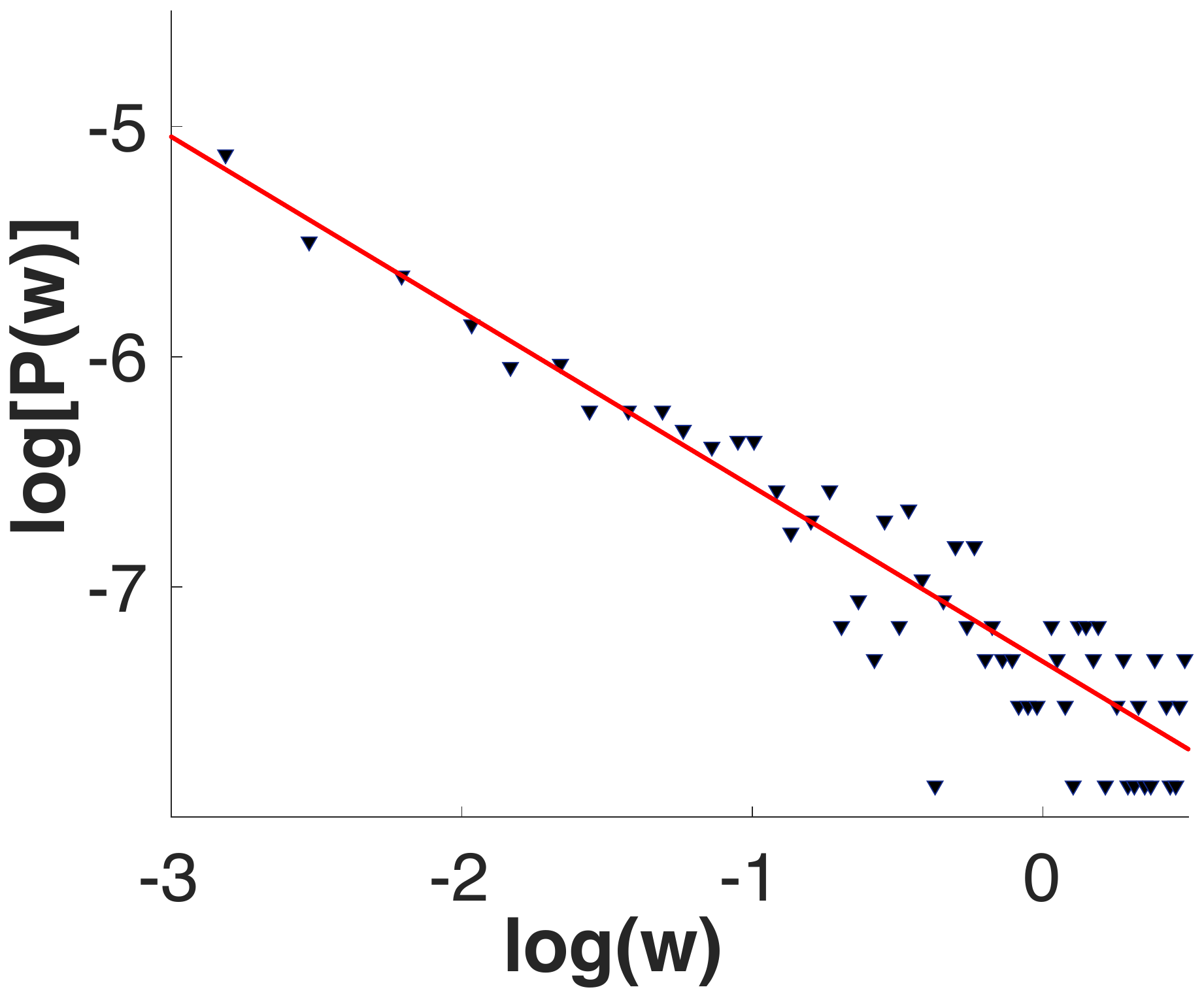}
b)\includegraphics[width=0.46\columnwidth]{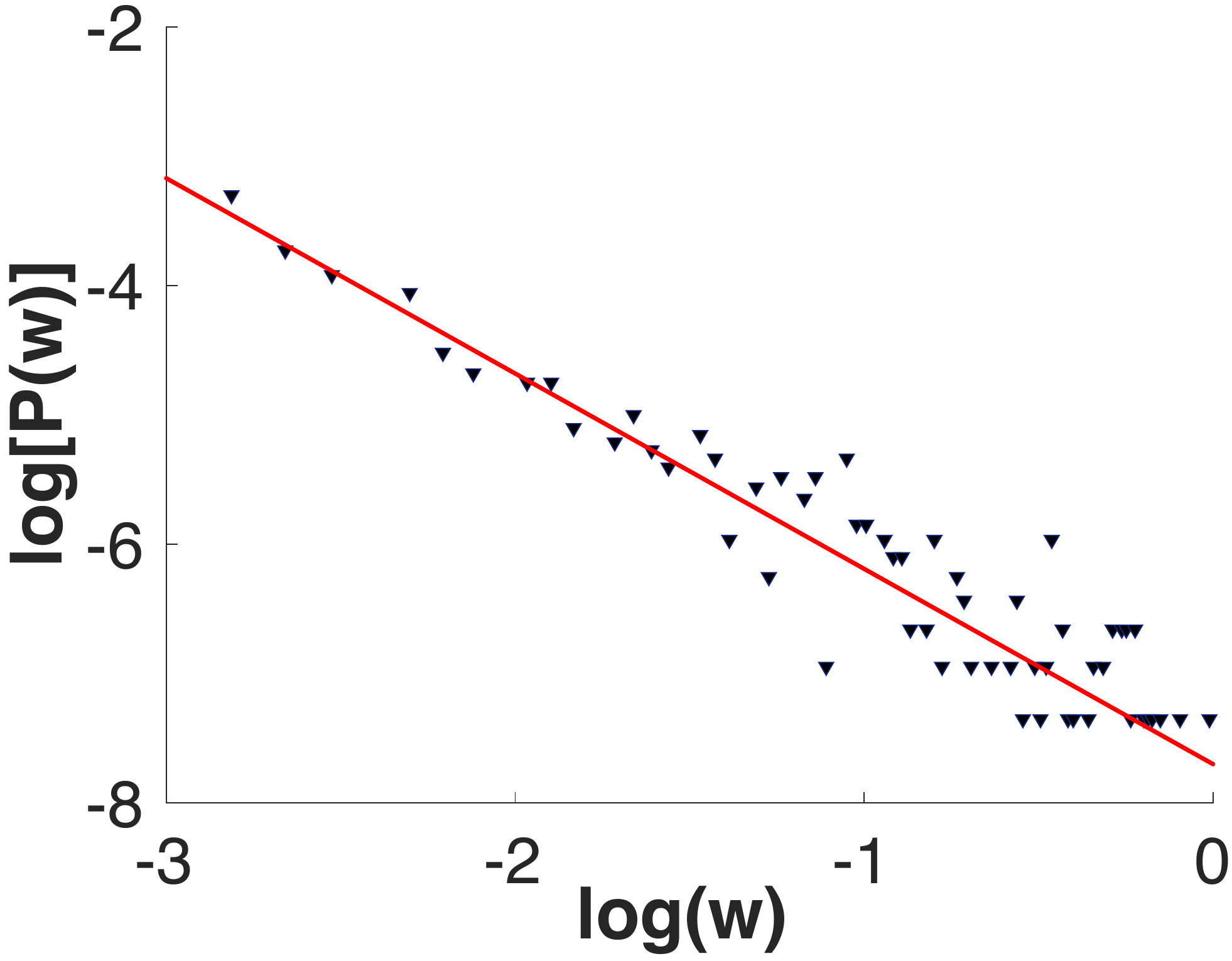}
\caption{Log-log plots for distributions of levels for a) $(\mu,\lambda)=(0.5,0.5)$, b) $(\mu,\lambda)=(1,2)$.}
\label{LS1}
\end{figure}

Fig.~\ref{LS1} shows the distribution of levels for $n=24$ and critical points $(\mu,\lambda)=(0.5,0.5)$ and $(\mu,\lambda)=(1,2)$.
These  can be fitted to the generic form: 
\begin{equation}
\log [P (w)]=-\delta\log(w)-\beta.
\end{equation}
For $(\mu,\lambda)=(0.5,0.5)$ and $(\mu,\lambda)=(1,2)$, $\delta = 1.52, 1.51$ respectively. We obtain a similar plot for   
$(\mu,\lambda)=(1,1)$, with slope $\delta = 1.52$. Thus along the critical lines the 
estimated value of $\delta$ is $\sim 1.5$, lending further  support to universality along the critical lines.

\section{Next nearest neighbour hopping}
In our study so far, we observe that \emph{all} the eigenfunctions  
in each of the phases exhibit features similar to the corresponding ground state eigenfunction. 
In other words, the nearest neighbor hopping model we considered is characterised by a lack of mobility edge.

Here we will study the role of next-nearest-neighbor hopping on the
localization-delocalization transition. For this purpose we will consider the following 
Harper equation~\cite{Beugeling}:
\begin{eqnarray}
E\Psi_n =D_n \Psi_n+ R_n \Psi_{n+1}+R_{n-1}^{\dagger} \Psi_{n-1}.
\end{eqnarray}
Here $D_n$ and $R_n$ are $2\times2$ matrices,
 \begin{equation}
D_n = \begin{pmatrix}
a_n &~ c_n \\ \\
c_n^{\dagger} &~b_n\\ \\
 \end{pmatrix}, ~~R_n = \begin{pmatrix}
d_n &~  0 \\ \\
f_n &~e_n\\ \\
 \end{pmatrix},   \notag
\end{equation}
where at $k_x=k_y=0$,  the matrix entries are given in terms of the NNN hopping parameter $t$ and the NN hopping parameters $t_1,t_2,t_3$ as:
\begin{align*}
a_n &= 2t \sin[2\pi(n -1/6) \phi ],\quad b_n = -2t\sin[2\pi(n +1/6) \phi ],\\ 
d_n &= -i t e^{i2\pi (n+1/3) \phi}  (1- e^{i2\pi(n+1 /3)\phi}),\\  
e_n &= i t  e^{i2\pi (n+2/3) \phi}  (1- e^{i2\pi(n+2 /3)\phi}),\\
c_n &= t_2 + t_1 e^{i 2\pi \phi n },\quad f_n = t_3 e^{-i 2\pi \phi (n+1/2)} .
\end{align*}

We note that our spinless  Harper equation is a modified version of the one originally  proposed by Beugeling et al.~\cite{Beugeling} for spinfull fermions with isotropic NN hopping and  the NNN hopping terms being of the intrinsic spin-orbit (ISO) and Rashba types, which correspond to spin conserving and non-conserving processes, respectively.  In the intrinsic spin-orbit term, the up and down-spin terms have opposite sign in front of the tunnelling amplitude term. We have considered only the ISO type of hopping process corresponding to spinless fermions.

\begin{figure}[H]
a)\includegraphics[width=0.46\columnwidth]{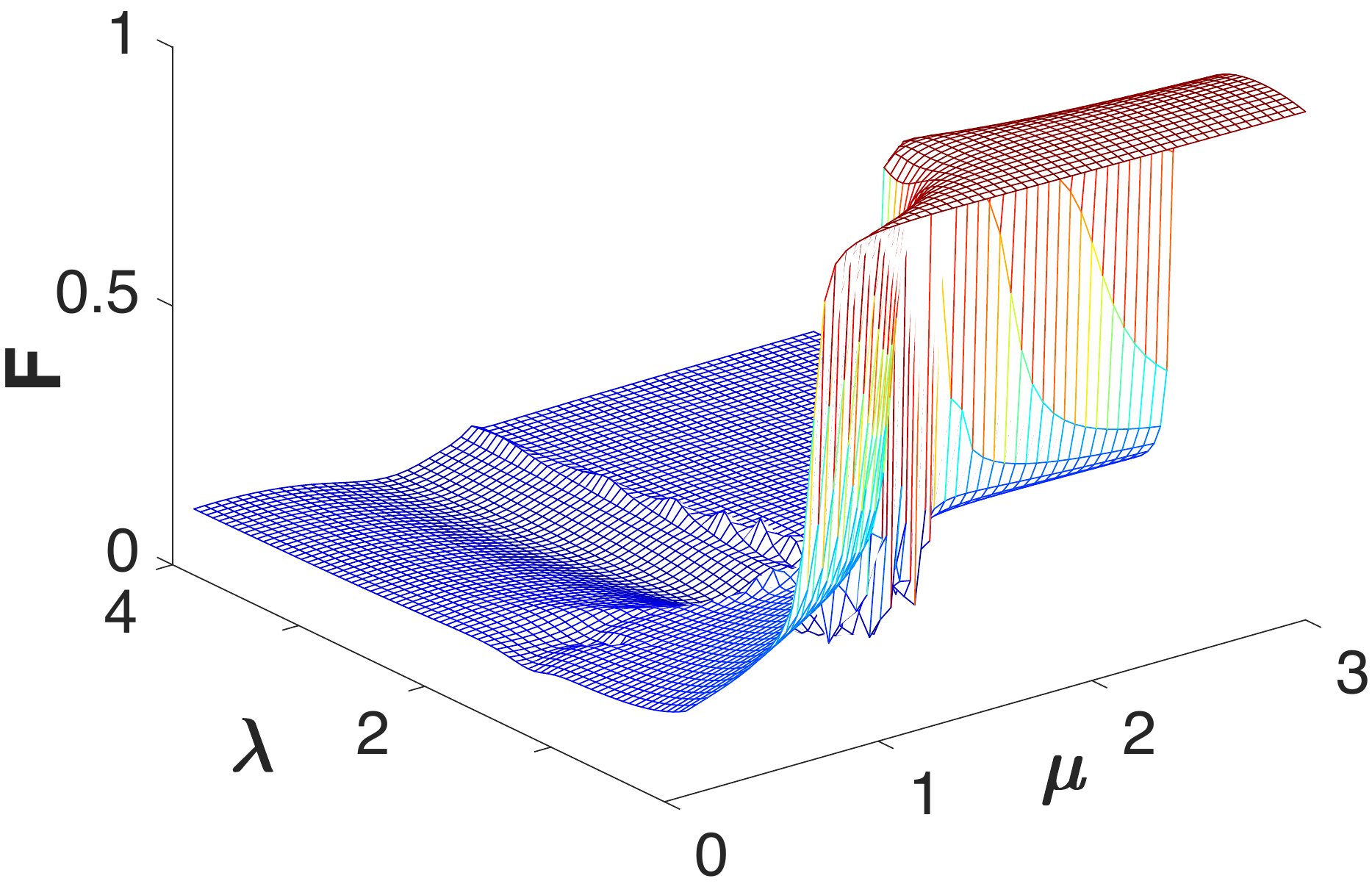}
b)\includegraphics[width=0.46\columnwidth]{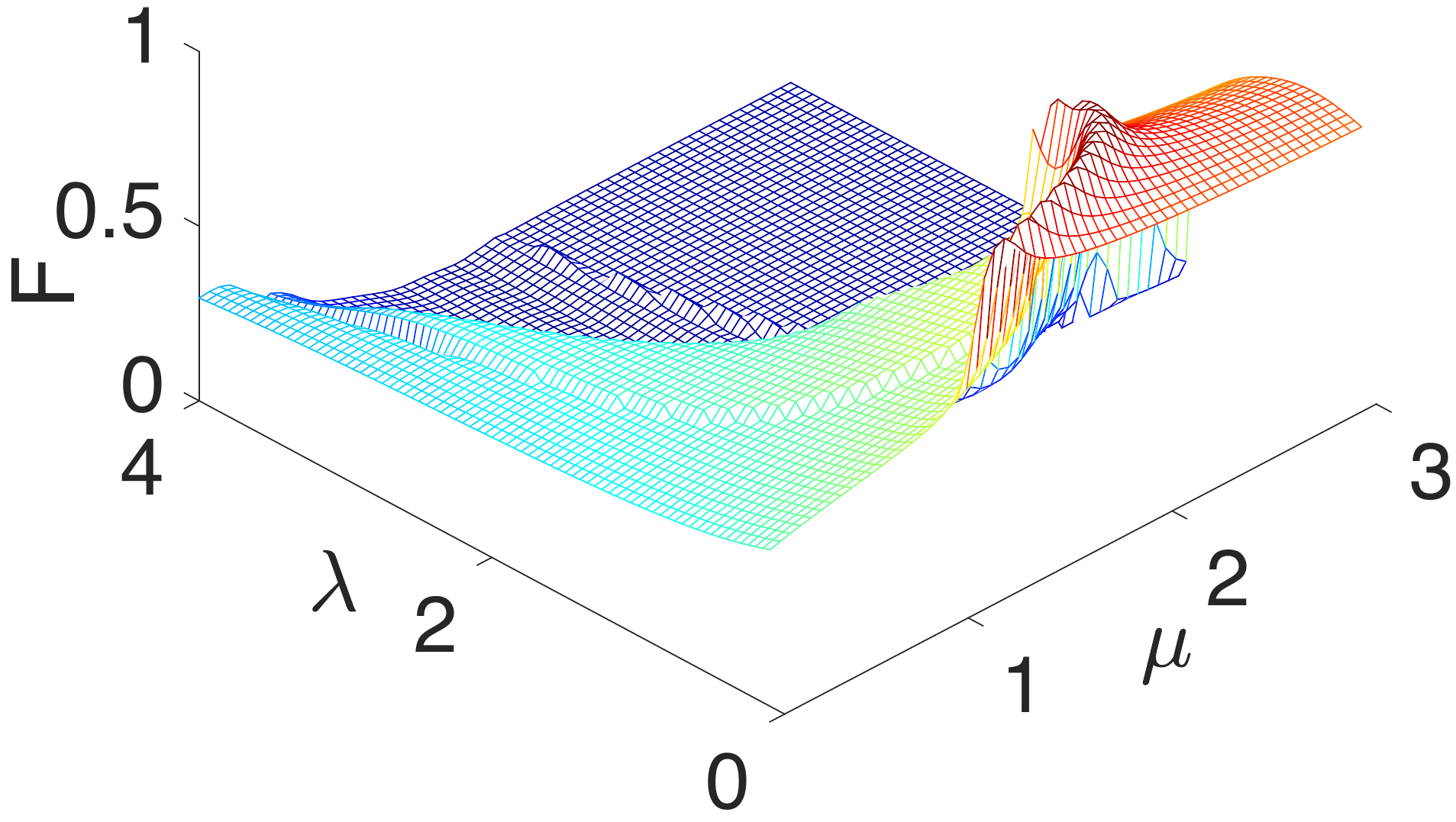}
\caption{Fidelity plots with ($\mu=2,\lambda=0.5$) and next nearest hopping terms a) $t=0.08$ and b) $t=0.12$.}\label{fidelityNNN}
\end{figure}

 The effective one dimensional model as before has asymmetric NN hopping; in addition,  the NNN hopping in the original hexagonal lattice manifests itself as an on site potential and NNN hopping along the chain. The Hamiltonian has the following form:
 \begin{widetext}
\begin{eqnarray}
H= \sum_n (a_n \alpha^\dagger_{2n} \alpha_{2n}  + b_n \alpha^\dagger_{2n+1} \alpha_{2n+1} )  + \sum_n (c_n^{*} \alpha^\dagger_{2n+1} \alpha_{2n}  + f_{n-1} \alpha^\dagger_{2n-1} \alpha_{2n} +d_n^{*} \alpha^\dagger_{2n+2} \alpha_{2n}  +
e_{n}^* \alpha^\dagger_{2n+3} \alpha_{2n+1} +h.c. ) .
\end{eqnarray}
\end{widetext}
In the absence of the NNN term we recover identically the earlier plots for the phase diagram, 
von Neumann entropy, fidelity and  fidelity-susceptibility. However,  turning on the NNN term leads to features that are not universal.    
 In the absence of $t$, the fidelity plots for reference parameters $(\mu=2,\lambda=0.5)$ will reproduce  Fig.~\ref{fig_fidelity} (b), however,
  when $t=0.08,\text{and~}0.12$,  the ground state fidelity yields plots as given in  Fig.~\ref{fidelityNNN} (a) and (b), respectively. These plots
   are  significantly different  from each other as well as from the $t=0$ case, indicating strong dependence of the nature of the ground state on
    the NNN hopping parameter. 
    
\begin{figure}[H]
a)\includegraphics[width=0.46\columnwidth]{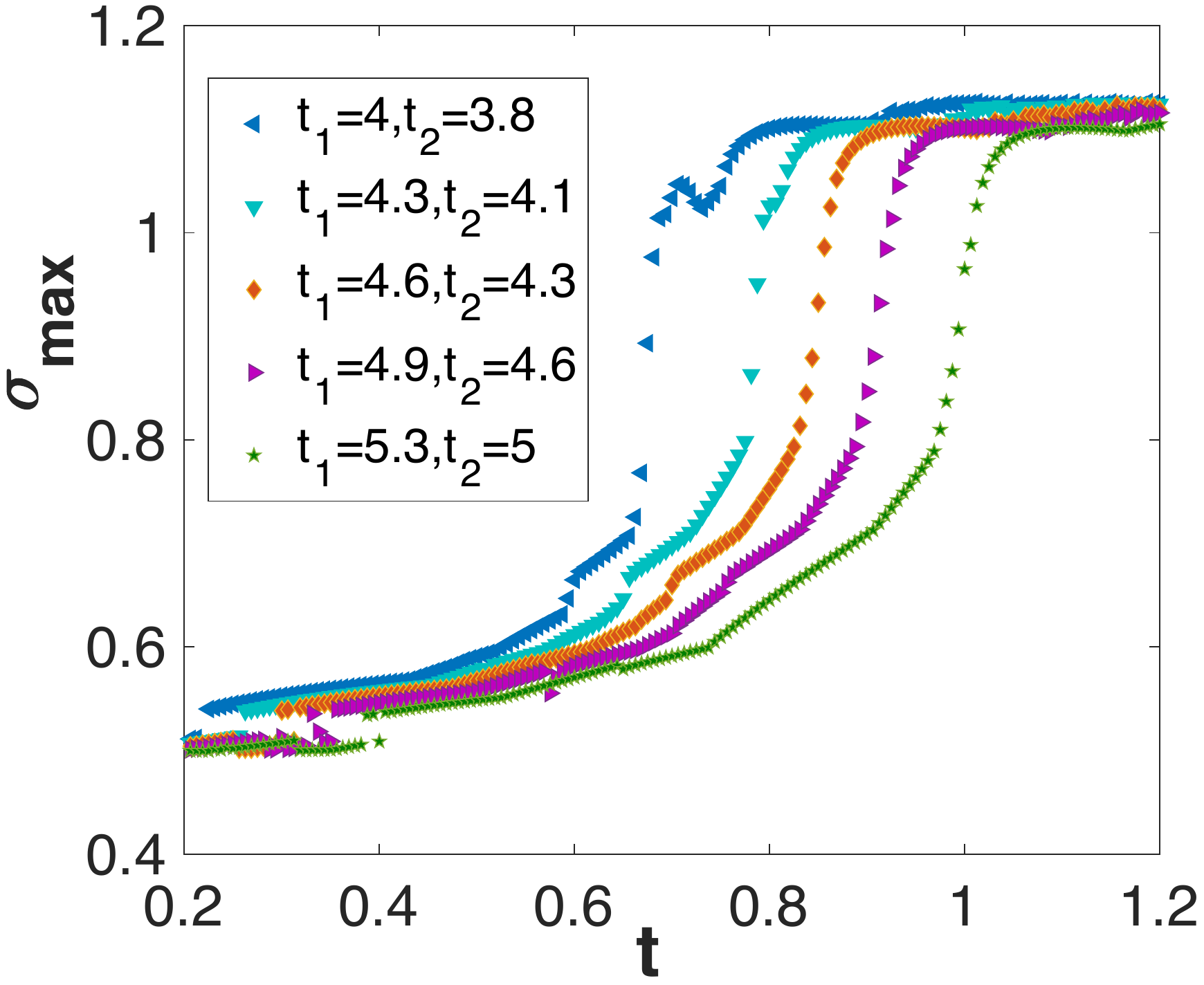}
b)\includegraphics[width=0.46\columnwidth]{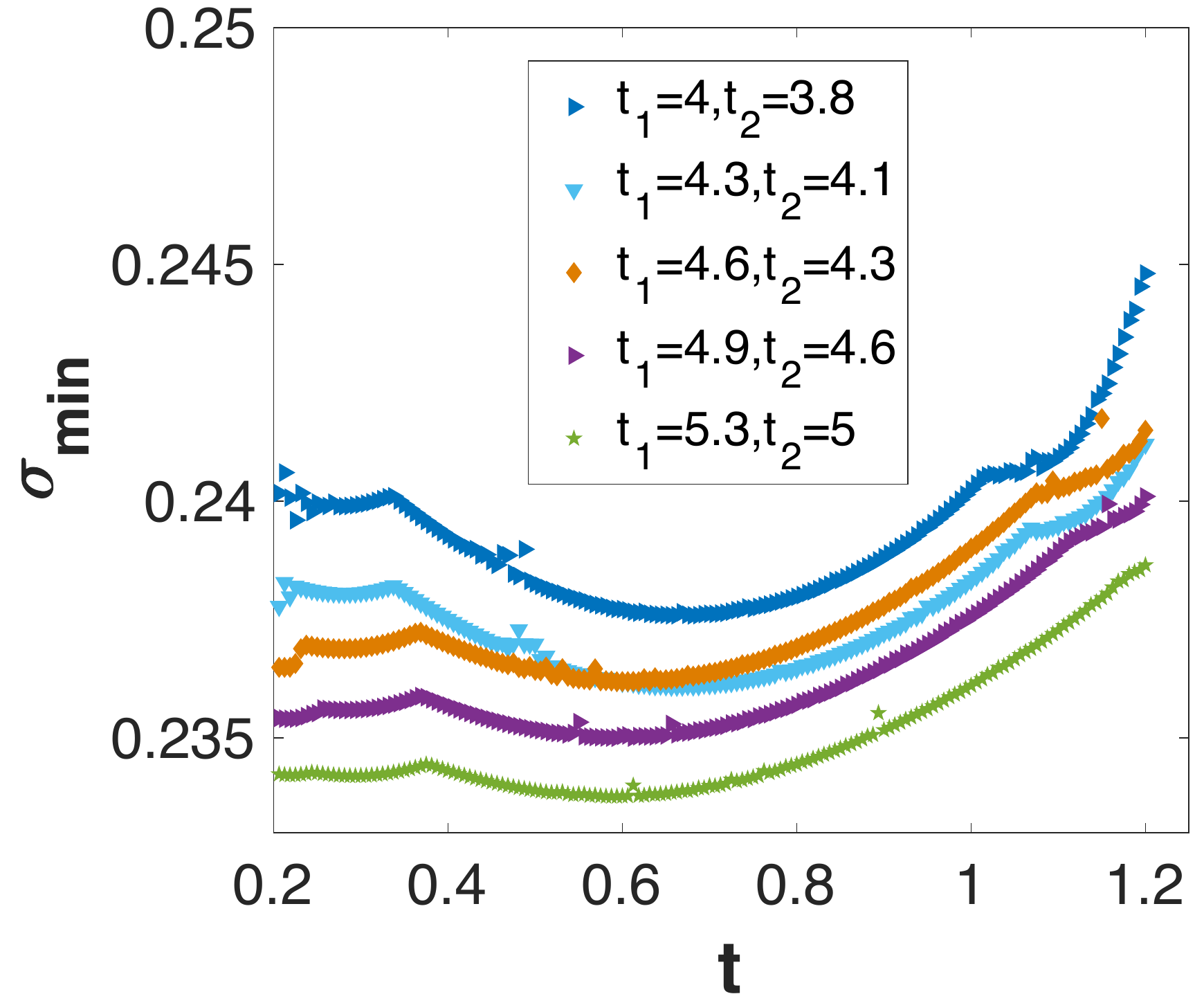}
\caption{Scaled entropy (Eq.~\ref{eqn:sigma}) plots for the NNN model for a range of  parameters $(t_1, t_2)$ as a function of the 
NNN coupling $t$.  a) shows the maximum entropy among all the eigenstates. Suggests a transition from localized to delocalized nature. 
b) shows the minimum entropy, which do not indicate a transition in nature, within the given range.}\label{S_max_vs_t}
\end{figure}
It turns out that unlike the NN model, the ground state here is \emph{atypical}; there are other eigenstates which are dramatically
 different from the ground state. To show this, we once again study the maximum and minimum values of the scaled entropy, 
 just like in Fig.~\ref{fig_entropy} for the NN problem, where $\sigma_{min}$ and $\sigma_{max}$ were qualitatively 
 the same. In sharp contrast, Fig.~\ref{S_max_vs_t} shows that 
in the presence of NNN coupling, the dependence as a function of $t$ of $\sigma_{min}$ and $\sigma_{max}$ are markedly different. 
As $t$ is increased  there exists a critical $t_c$,  dependent on the parameters  $(\mu,\lambda)$, beyond which 
the states corresponding to the maximum entropy exhibit a localisation-delocalisation transition (for large $n$ the transition is sharper). For large $t$, $S_{\text{max}} \sim \log_2 N/N$ implying some of the states become delocalised. There is no such signature seen with $\sigma_{min}$. This is an indication of the presence of mobility edges in the spectrum.

\begin{figure}
a)\includegraphics[width=0.46\columnwidth]{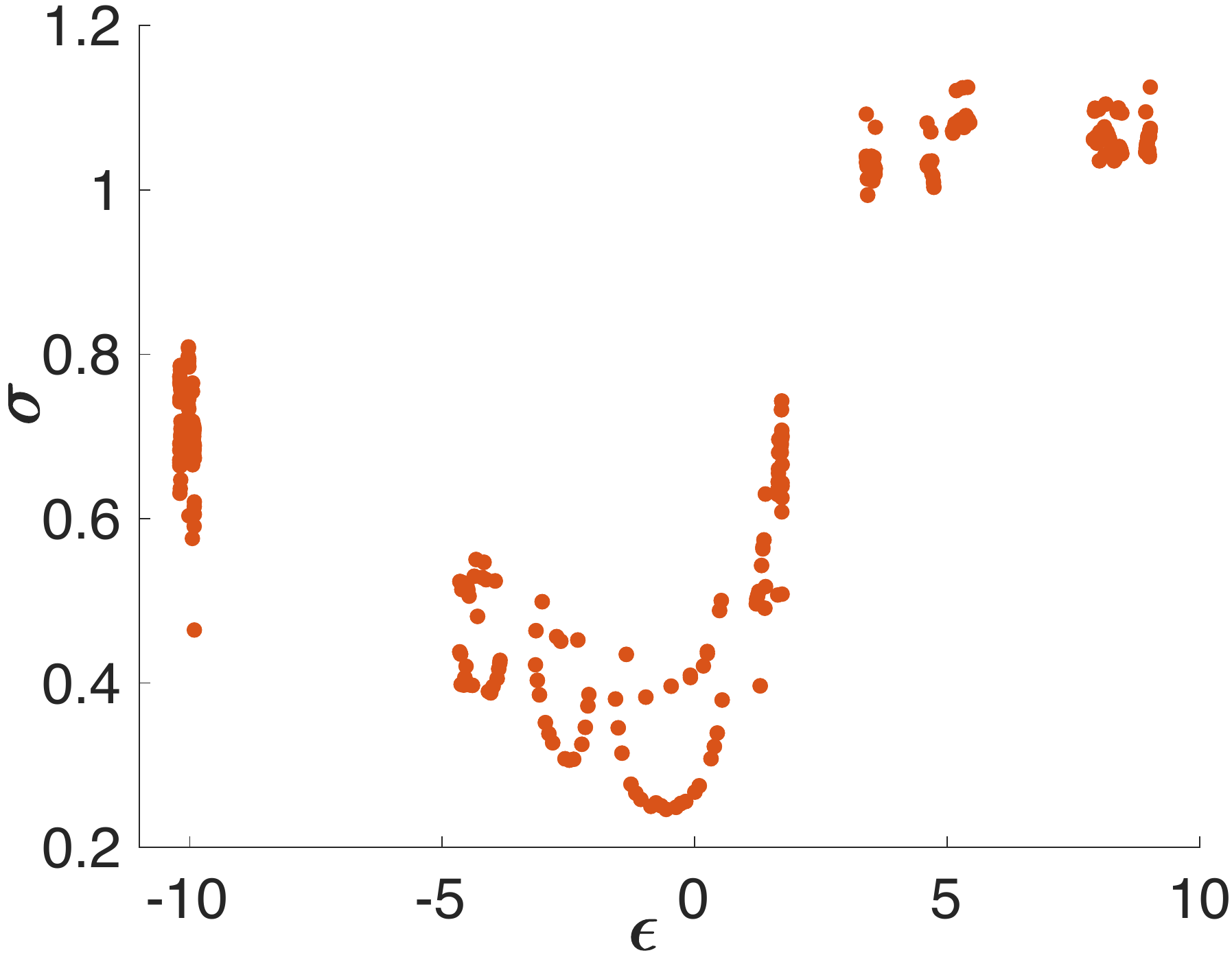}
b)\includegraphics[width=0.46\columnwidth]{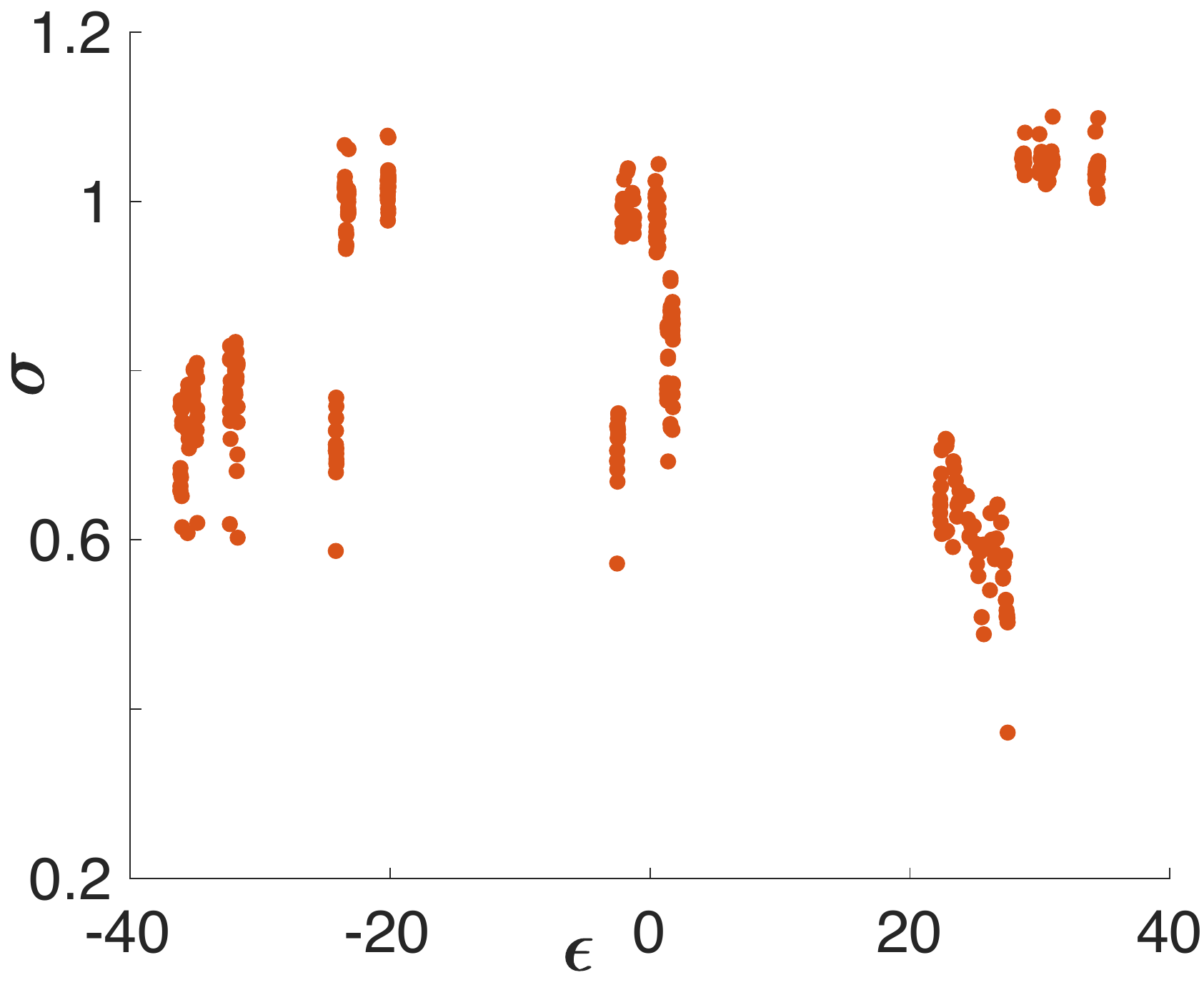}
\caption{Scaled entropy (Eq.~\ref{eqn:sigma}) plots for all energies of the NNN model with a) $t=1.2$ and b) $t=10$. }\label{S_max_epsilon}
\end{figure}

A closer look at the plateau region of Fig.~\ref{S_max_vs_t} a) is illuminating. Fixing $t$ to be in the plateau region (and $t_1=4, t_2=3.8$), let us study the scaled entropy for all the eigenstates as a function of energy. Two different cases of $t = 1.2$, and $t=10$ are shown in Fig.~\ref{S_max_epsilon}. In each of these cases, it is possible to discern  spread-out energy windows where $\sigma \sim 1$ corresponding to extended states, and other energy windows where $\sigma \ll 1$ corresponding to localized states. 
For $t=1.2$  a mobility edge  separating localized and delocalized states can be identified to be around $\epsilon \approx 3$. However, such sequential behaviour is not universal as evidenced for $t=10$, wherein the behaviour  exhibited is more complex.  There are multiple closely separated energy  bands at various energy scales wherein the wavefunctions manifest different behaviour: delocalized, localized and critical.

\section{Conclusions}
The phase diagram of the 2D honeycomb lattice with asymmetric nearest neighbour hopping in the presence of a uniform magnetic field is obtained, employing a variety of techniques. It is found to comprise of one insulating region and two metallic regions; von Neumann entropy, averaged over all sites, is useful as a marker of the phase boundaries. The task of distinguishing between the two metallic phases, is accomplished  with the help of fidelity, and fidelity susceptibility. An alternative perspective is provided by multifractal analysis, which allows for a finite-size scaling confirmation of the phase diagram. The study of level statistics reveals universal features along the critical lines. Finally, a variant Hamiltonian which incorporates next nearest neighbour hopping, is found to exhibit mobility edges in contrast to the earlier model.

While the  present work has focussed  on statics,  a dynamical perspective is suggested by recent work~\cite{morales} which has reported on certain driving protocols that can couple between groups of localized states and convert them into extended states. It would be interesting to explore such effects in our model on the honeycomb lattice, which provides a rich phase diagram. The possibility of many-body localization when interaction is turned on~\cite{modak}, and other many-particle effects~\cite{Wang}, provides impetus for including interaction in our model.  
The AAH model has been shown to be  topologically nontrivial~\cite{Kraus}; it remains to be seen whether our 1D  chain admits a topological phase. 
\acknowledgments
S.G. would like to thank Diptiman Sen for helpful discussions. A.S  and S.G. are grateful to  SERB for the support via grants  YSS/2015/001696
and EMR/2016/00264, respectively. We thank the HPC facility of IISER, Bhopal, where part of the computational work  was carried out.


\end{document}